\documentclass[preprint,showpacs,preprintnumbers,amsmath,amssymb]{revtex4}

\usepackage{graphicx}
\usepackage{dcolumn}
\usepackage{bm}

\begin{document}

\title{
Collapse of the hyperfine magnetic field at the Ru site in 
ferromagnetic rare earth intermetallics}

\author{D. Coffey} 

\affiliation
{Dept. of Physics, Buffalo State College, Buffalo, New York 14222
}

\author{M. DeMarco}

\affiliation
{{Dept. of Physics, Buffalo State College, Buffalo, New York 14222}
 and {Dept. of Physics, SUNY Buffalo, Buffalo, NY 14260}}

\author{P. C. Ho}

\affiliation
{Dept. of Physics, California State, Fresno CA 93740}

\author
{M. B. Maple and  T. Sayles}

\affiliation
{Dept. of Physics, University of California, San Diego CA 92093}

\author
{J. W. Lynn and Q. Huang}

\affiliation
{NIST Center for Neutron Research, National Institute of Standards and Technology, Gaithersburg, MD 20899}

\author
{S. Toorongian and M. Haka}

\affiliation
{Nuclear Medicine Department, State University of New York, NY 14260.}

\begin{abstract}
The M\"{o}ssbauer Effect(ME) is frequently used to investigate magnetically ordered systems. 
One usually assumes that the magnetic order induces a hyperfine
magnetic field, $B_{hyperfine}$, at the ME active site.
This is the case in the ruthenates, where the temperature dependence
of $B_{hyperfine}$ at $^{99}$Ru sites tracks the temperature dependence 
of the ferromagnetic or antiferromagnetic order.
However this does not happen in the rare-earth intermetallics, GdRu$_2$ and HoRu$_2$.
Specific heat, magnetization, magnetic susceptibility, M\"{o}ssbauer effect, and neutron diffraction 
have been used to study the nature of the magnetic order in these materials.  
Both materials are found to order 
ferromagnetically at 82.3 and 15.3 K, respectively.
Despite the ferromagnetic order of the rare earth moments in both systems,
there is no evidence of a correspondingly large
 $B_{hyperfine}$ in the M\"{o}ssbauer spectrum at the Ru site. 
Instead the measured spectra consist of  a narrow peak  at all temperatures 
which points to the absence of magnetic order.
To understand the surprising absence of a transferred hyperfine magnetic field,
we carried out {\it ab initio} 
calculations which show that 
spin polarization is present only on the rare-earth site.
The electron spin at the Ru sites is effectively unpolarized 
and, as a result, $B_{hyperfine}$ is very small   at those sites.
This occurs because the 4$d$ Ru electrons form broad conduction bands 
rather than localized moments.
These 4$d$ conduction bands are polarized in the region of the Fermi energy and
mediate the interaction between the localized rare earth moments.
\end {abstract}

\date{\today}

\maketitle

\section{Introduction}
\vskip 8pt
Compton and Matthias\cite{Compton1959} found that 
superconductivity and ferromagnetism occur in
Laves phase compounds containing
lanthanide
elements and Ru.
This led to the suggestion that superconductivity and ferromagnetism 
could coexist in Ce$_{1-x}$R$_x$Ru$_2$ alloys, since superconductivity occurs below 6 K in CeRu$_2$
and ferromagnetism is present in that temperature range in RRu$_2$, where R is
Gd or Ho\cite{Matthias1958}.
The phase diagram for  Ce$_{1-x}$R$_x$Ru$_2$
was investigated by Wilhelm and Hillebrand\cite{Wilhelm1971} for R=Gd, Ho, Dy, Pr.
Although no long range magnetic order was found in the superconducting region
of the phase diagrams of these alloys, evidence for
short-range order was found in the temperature dependence of the $^{155}$Gd ME 
and in nuclear quadrupole resonance
measurements. This occurs  in Ce$_{1-x}$Gd$_x$Ru$_2$\cite{Ruebenbauer1977,Kumagai1978} 
for a narrow range of doping about $x\sim 0.1$.
Fischer and Peter\cite{Fischer1973} pointed out that the specific heat of 
Ce$_{1-x}$Gd$_x$Ru$_2$ measured by Peter et al.\cite{Peter1971} also showed an
anomalous temperature dependence.
The specific heat divided by temperature, $\frac{C}{T}$, for CeRu$_2$ 
showed a sharp jump at the superconducting transition and a rapid fall off to zero as $T \rightarrow 0$, as expected.
However the jump becomes more rounded in Ce$_{1-x}$Gd$_x$Ru$_2$ as $x$ increases and
$\frac{C}{T}$ increases as $T \rightarrow 0$.
The value of $\frac{C}{T}$ as  $T \rightarrow 0$ increases as $x$ increases from 0.05 to 0.11.
This anomalous temperature dependence was taken as evidence of a ferromagnetic contribution 
to $\frac{C}{T}$.  The analysis of the data did not provide a microscopic model for the nature of this 
contribution or explain how it could coexist with superconductivity.

Evidence for the coexistence of superconductivity and short-range ferromagnetic correlations 
was also found
in Ce$_{0.73}$Ho$_{0.27}$Ru$_2$ from the temperature dependence of the $^{57}$Fe ME
below $\sim$2 K and 
from neutron scattering data\cite{Lynn1979,Lynn1980,Willis1980}.
Since the hyperfine coupling constant of Ru is twice that of Fe\cite{Watson1977} it was expected
that there would be a magnetic field at the Ru nucleus of about 15 T in this material.
Hyperfine magnetic fields were reported at the Gd site in Ce$_x$Gd$_{1-x}$Ru$_2$\cite{Ruebenbauer1977}
 for $x > 0.2$, whose temperature dependence was consistent with the Curie temperature.
More recently, Andoh discussed the magnetic properties of a number of the RRu$_2$\cite{Andoh1987}.
\vskip 8pt
CeRu$_2$ has been assigned to have the cubic Laves structure(Fd$\bar{3}$m), 
although recently Huxley et al.\cite{Huxley1997} have shown that 
its symmetry is lowered due to a slight variation in
the displacement of the Ru from their cubic Laves positions.
GdRu$_2$ and HoRu$_2$ are in  the hexagonal Laves phase
structure (P6$_3$/mmc). GdRu$_2$ has also been reported in the cubic Laves phase.\cite{Krip1963}
The difference in structure does not determine whether the ground state is
superconducting or ferromagnetic since NdRu$_2$, a ferromagnet, has the Fd$\bar{3}$m structure.
Here we concentrate on reconciling the ferromagnetism found 
in neutron scattering, transport, and thermodynamic measurements 
with M\"{o}ssbauer spectroscopy.
\vskip 8pt
M\"{o}ssbauer spectroscopy is a  nuclear probe of the electronic 
properties of systems which has been used to investigate 
magnetic order in many systems.  
The evidence of magnetic order appears in the M\"{o}ssbauer spectrum as a hyperfine
magnetic field induced at the nucleus at which the M\"{o}ssbauer Effect(ME) is measured.
In this way the temperature dependence of magnetic order 
has been probed
by the $^{99}$Ru ME in both ferromagnetic and 
antiferromagnetic ruthenates\cite{DeMarco2000,Coffey2008}.
The magnetic order has also been investigated using the $^{57}$Fe,
$^{193}$Ir, and $^{155}$Gd ME
in the intermetallic compounds of interest here.
One can distinguish between two cases. In the first case
the $B_{hyperfine}$ is found at a site 
on which there is an ordered electronic moment. 
This is the case of the ruthenates and a number of rare earth intermetallics.
Using the $^{57}$Fe ME, Wertheim and Wernick\cite{Wertheim1962}
measured $B_{hyperfine}$ values in RFe$_2$(R=Ce, Sm, Gd, Dy, Ho, Er, Tm).
$B_{hyperfine}$ 
is found to be  $\sim$23T in spite of the wide range in the size of the localized moments 
on the R sites. By comparison, the value for $B_{hyperfine}$ in ferromagnetic Fe is 32 T.
The $^{57}$Fe ME has also been used to investigate the magnetic structure in 
 rare earth iron ternary intermetallics.\cite{Atzmony1973}
De Graaf et al.\cite{Graaf1982} extracted a value for $B_{hyperfine}$ equal to 17.5T at the 
Gd site in GdCu$_2$ using $^{155}$Gd ME from a structureless spectrum.
\vskip 8pt 
In the second case,  for a non-magnetic ion in a magnetically ordered lattice, the measured
ME at this site is expected to show evidence of the magnetic order through a transferred hyperfine magnetic field.
Transferred hyperfine fields at the non-magnetic Ir site were measured by Atzmony et al.\cite{Atzmony1967}
in 
RIr$_2$(R=Pr,Nd,Sm,Gd,Tb,Dy,Ho). They  found a wide variation from 4 T in HoIr$_2$ 
to 19 T in GdIr$_2$.
Transferred hyperfine fields have also been measured in a rare-earth
matrix doped with 1$\%$ Sn and in R$_2$Sn using the $^{119}$Sn ME\cite{Bosch1966}. For Sn doped into a rare earth,
these range from -5.3 T 
in a Tm matrix to 23.8 T in a Gd matrix. The values of $B_{hyperfine}$ are 
linear in the projection of the spin of the rare earth moment on its total angular momentum,
$(g-1)J$.
In R$_2$Sn, these fields range from -5.5 T (Er$_2$Sn) to 28.9 T (Gd$_2$Sn).
\vskip 8pt
The fact that the sign of the transferred hyperfine field can change suggested that there is competition between 
different contributions which align or antialign the nuclear moment with the ordered electronic moment.
A negative hyperfine magnetic field is antiparallel to the ordered electronic moment.
Watson and Freeman were among the first to investigate the origin of the hyperfine magnetic field with
large scale numerical calculations based on the Hartree-Fock approximation\cite{Watson1961a,Watson1961b}.
Although these calculations were limited by the computational capabilities then available, they demonstrated
a number of the qualitative features of the experimental data.
In particular, they showed that the ordered $4f$ moment in Gd polarizes the electron density 
in the opposite direction to the ordered $4f$ moment very close to the nucleus
  and  in the region beyond  
localized $4f$ bands at the edge of the Gd ion. 
They showed that the largest contribution to the hyperfine magnetic field due to polarization of the spin density 
of $s$ electrons is the result of contributions of different signs from different $s$-shells.
Their calculations also  showed that the polarization of the spin density on neighboring 
sites 
could be opposite to that of the ordered
moments leading to the negative hyperfine magnetic field.
The sign of the transferred hyperfine field can be modulated by varying the lattice constant, as we describe in the section on 
calculations of the electronic properties below.
In the calculations of Watson and coworkers, it was assumed that the $5s$ electrons formed the conduction band
which turns out not to be the case in GdRu$_2$, as we will also discuss below.
In addition 
orbital contributions to transferred hyperfine magnetic field, based on the assumption that 
it arises from the coupling of $f$ electrons on the rare earth sites with $s$-conduction electrons,
were investigated by
Dunlap et al.\cite{Dunlap1973}
Local Spin Density Approximation(LSDA)
 calculations were first used to calculate hyperfine magnetic fields 
in ferromagnetic $3d$ metals by Callaway and Wang\cite{Callaway1977,Wang1977}.
We use a more recent implementation of the LSDA to calculate hyperfine fields on 
Ru sites and compare with our measured values.
\vskip 8pt 
We present magnetic, transport and thermodynamic data on GdRu$_2$, and magnetic and neutron diffraction
data on HoRu$_2$, showing that these are ferromagnetically ordered at low temperatures.
However, our ME measurements of 
GdRu$_2$ and HoRu$_2$ show that $B_{hyperfine}$ at the Ru site is 
so small that, without the evidence of other experiments, one would conclude that there is no magnetic order.
The almost complete collapse of the value of $B_{hyperfine}$
is an unexpected result in materials whose Curie temperatures are 82.3 K and 15.3 K,
respectively.  Interestingly, the absence of $B_{hyperfine}$
at the Ru sites in GdRu$_2$ was noted previously without comment\cite{Kistner1966}.

We calculate the electronic 
properties of these materials using a spin polarized fully relativistic all-electron
linearized augmented plane wave method\cite{wien2k}.
{\it ab initio} bandstructure calculations have previously been used by other 
authors to determine hyperfine magnetic fields and electric 
field gradients (EFG).\cite{Lippens2000,Blaha2000,Cavor2004,Pav2006}
We find good agreement between the EFG tensor 
and that found from the M\"{o}ssbauer 
spectra of CeRu$_2$.
We also find that the calculated $B_{hyperfine}$ at the Ru sites in GdRu$_2$ and HoRu$_2$ are much smaller than
those on Gd and Ho, consistent with the experimental results.
First we present the experimental results.

\section{Experimental Results }

\subsection{ Evidence of Ferromagnetism}
\begin{figure}
 \begin{center}
 \includegraphics[angle=270,width=1\textwidth]{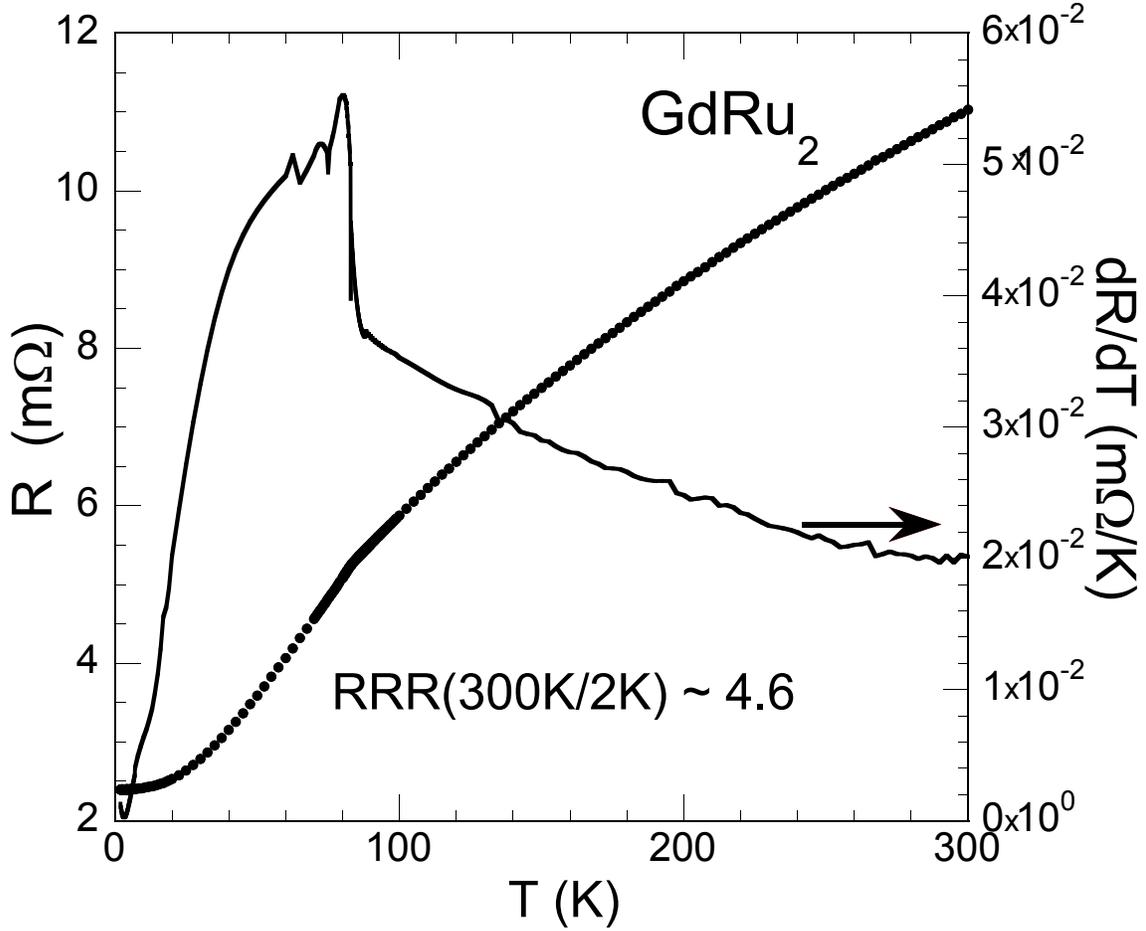}
 \caption{Temperature $T$ dependence of resistance $R$ and the slope $dR/dT$ of
          a polycrystalline sample of GdRu$_2$. The value of the residual resistivity
          ratio RRR(300 K/2 K) is $\sim 4.6$.  A curvature breaking which occurs in
          the $R(T)$ data at $\sim 80$\,K corresponds to the transition temperature of magnetic order.}
 \label{RvsTGdRu2}
 \end{center}
\end{figure}

\begin{figure}
 \begin{center}
 \includegraphics[angle=270,width=1\textwidth] {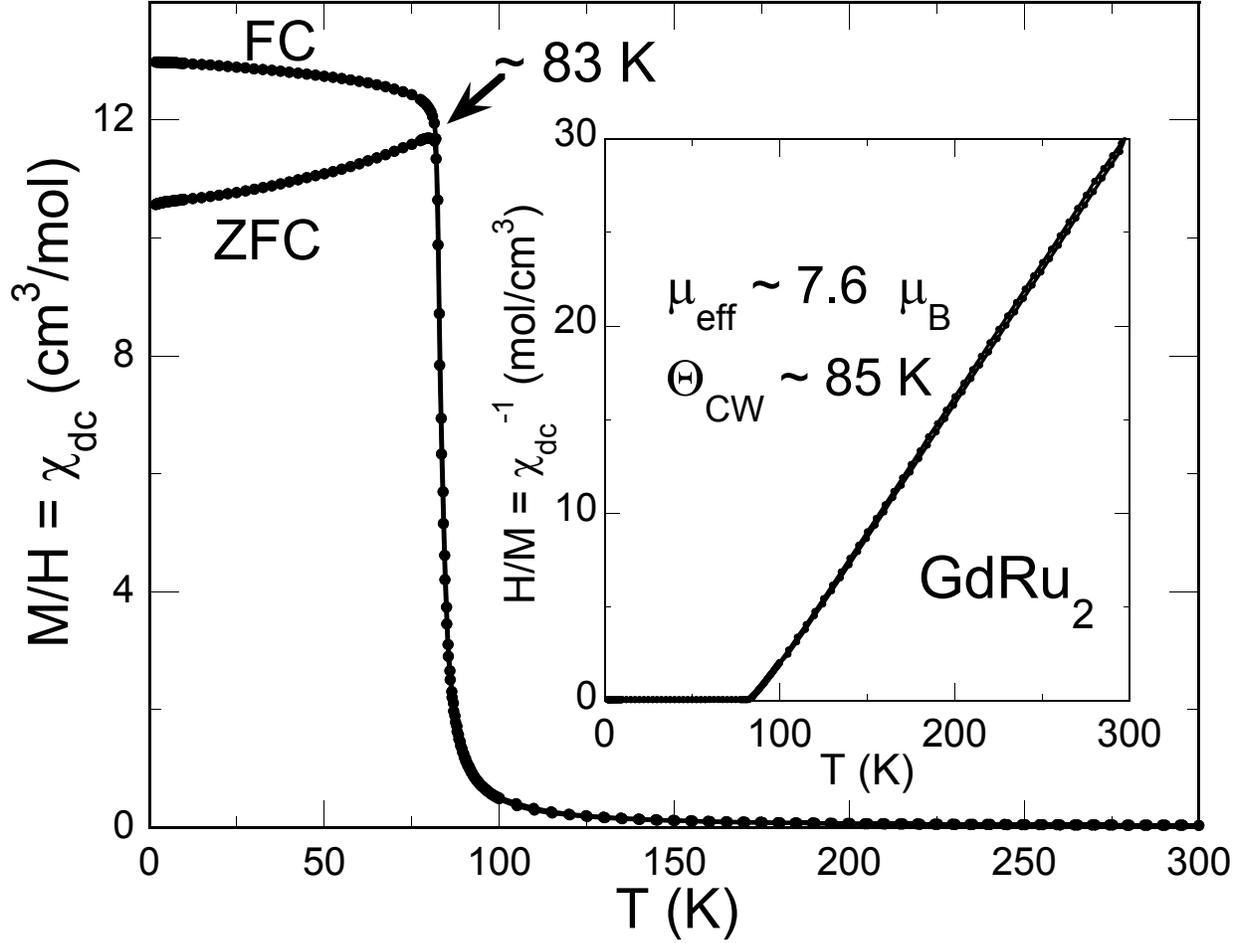}
 \caption{Temperature dependence of dc-magnetic susceptibility $\chi_{\rm{dc}}$ of GdRu$_2$
          measured at an applied magnetic field $H = 50$\,Oe. Hysteresis occurs at $T \sim 83$\,K,
          coinciding with the Curie temperature T$_{\rm{C}}$ determined from the modified Arrott 
           plot\cite{Arrott1967}.
          $\chi_{\rm{dc}}^{-1}$ vs $T$ is shown in the inset and follows a Curie-Weiss behavior, 
           which results in
          an effective moment $\mu_{\rm{eff}} \sim 7.6 \mu_{\rm{B}}$ and
          a Curie-Weiss temperature $\Theta_{\rm{CW}} \sim 85$\,K.}
 \label{chidcvsTGdRu2}
 \end{center}
\end{figure}

The transport and thermodynamic properties of GdRu$_2$ were investigated with a number of probes.
The temperature, $T$, dependence of electrical resistance, $R$, and the
slope d$R$/d$T$ of a polycrystalline sample of GdRu$_2$ are plotted
in Fig.~\ref{RvsTGdRu2}. A breaking curvature in $R$, accompanied by
a sharp increase of d$R$/d$T$, occurs at $\sim 86$\,K as temperature
decreases, which is due to the development of an ordered state.
The dc-magnetic susceptibility
$\chi_{\rm{dc}}$ is measured from 1.9\,K to 300\,K at an applied
magnetic field $H = 50$\,Oe in the zero field cooled (ZFC) and field
cooled (FC) states and the data are displayed in
Fig.~\ref{chidcvsTGdRu2}. Hysteresis in $\chi_{\rm{dc}}(T)$
appears at $\sim 83$\,K. 
A Curie-Weiss analysis was done on the molar
magnetic susceptibility data $\chi_{\rm{mol}}$ given by
\begin{equation}
  \chi_{\rm{mol}}=
  \frac{1}{3}\frac{N_{\rm{A}}\mu_{\rm{eff}}^2}{k_{\rm{B}}(T-\Theta_{\rm{CW}})},
 \end{equation}
where $N_{\rm{A}}$ is the Avogadro's number,
$\mu_{\rm{eff}}=g(JLS)\sqrt{J(J+1)}\mu_B$ is the
effective magnetic moment,
$g(JLS)$ is the Land\'{e} $g$-factor,
$k_{\rm{B}}$ is the Boltzmann's constant,
and $\Theta_{\rm{CW}}$ is the Curie-Weiss temperature.  A positive
$\Theta_{\rm{CW}} \sim 85$\,K indicates a ferromagnetic order in
GdRu$_2$ and $\mu_{\rm{eff}} \approx 7.6$\,$\mu_B$ (Bohr magneton),
which is close to the theoretical value 7.94\,$\mu_B$ of the free
ion moment of Gd$^{3+}$. 
This value of  $\Theta_{\rm{CW}}$ is consistent with the temperature at which hysteresis first appears.

\begin{figure}
 \begin{center}
 \includegraphics[angle=0,width=0.8\textwidth]{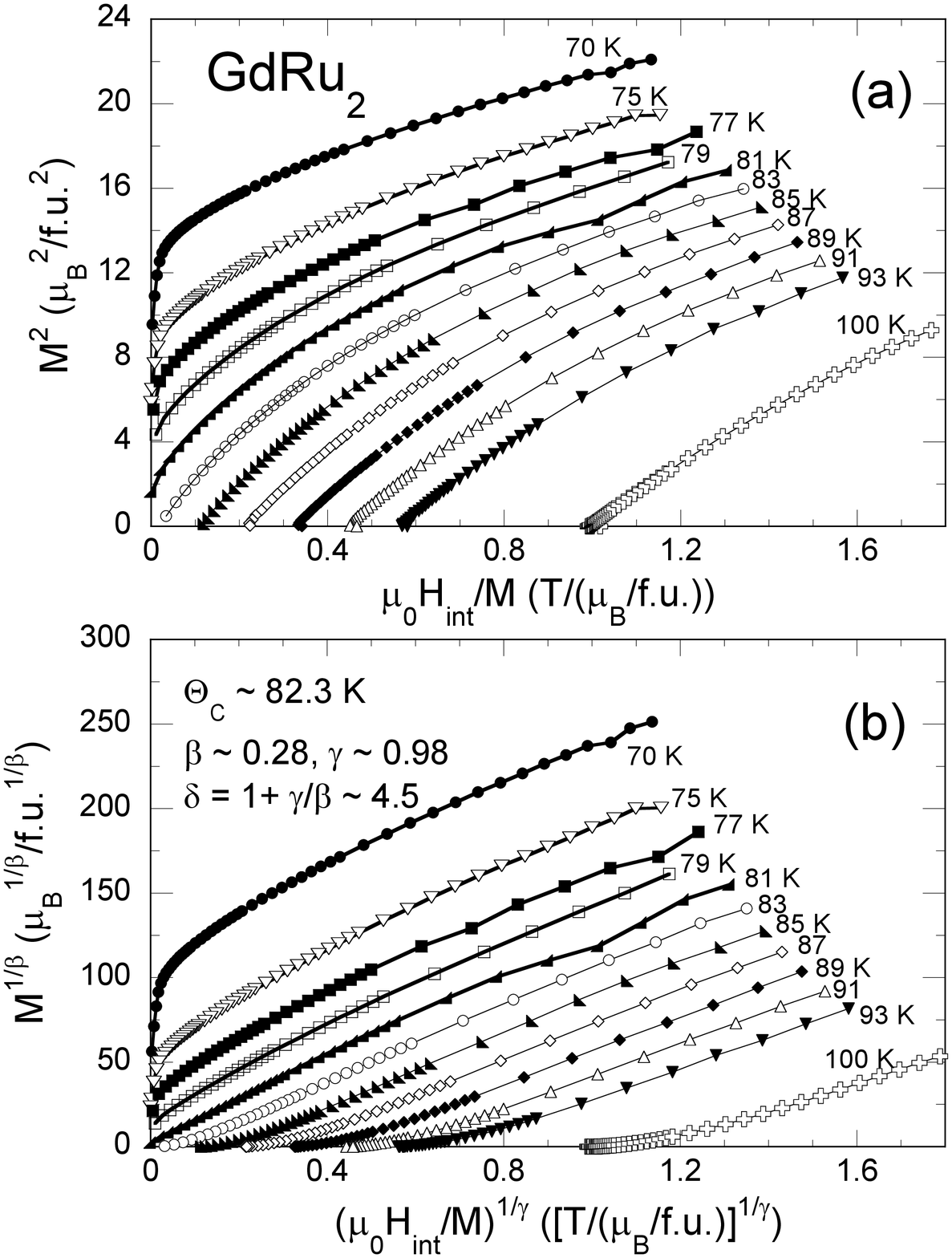}
 \caption{(a) Conventional and (b) modified Arrott plots of GdRu$_2$.
          The Curie temperature T$_{\rm{C}}$ of GdRu$_2$ is $\sim 82.3$\,K.
          The choice of exponents in (b) gives a closer to linear behavior in the critical region for intermediate fields.
          }
 \label{ArrottPlotsGdRu2}
 \end{center}
\end{figure}

Arrott plots of magnetization $M$ with respect to the internal
magnetic flux density $\mu_0 H_{\rm{int}}$ divided by $M$ were
constructed in an attempt to determine the Curie temperature
T$_{\rm{C}}$ more accurately. A conventional Arrott plot
consisting of $M^2$ vs \mbox{($\mu_0 H_{\rm{int}}/M)$} isotherms, is
shown in Fig.~\ref{ArrottPlotsGdRu2}(a).  
In the simplest mean-field analysis of 
ferromagnetism, $M^2$ vs \mbox{($\mu_0 H_{\rm{int}}/M$)} isotherms
form a series of parallel straight lines near T$_{\rm{C}}$,
and the isotherm passing through the origin corresponds to
T$_{\rm{C}}$.  However, the $M^2$-\mbox{($\mu_0 H_{\rm{int}}/M$)}
isotherms of GdRu$_2$ are slightly curved. Therefore, we applied a
modified Arrott plot $M^{1/\beta}$ vs
\mbox{($\mu_0H_{int}/M)^{1/\gamma}$}, where $\beta \sim 0.28$ and
$\gamma \sim 0.98$ are the critical exponents based on the
Arrott-Noakes equation~\cite{Arrott1967}, and the results
are plotted in Fig.~\ref{ArrottPlotsGdRu2}(b). The value
of T$_{\rm{C}}$ is $\sim 82.3$\,K.
\begin{figure}
 \begin{center}
 \includegraphics[angle=0,width=0.8\textwidth]{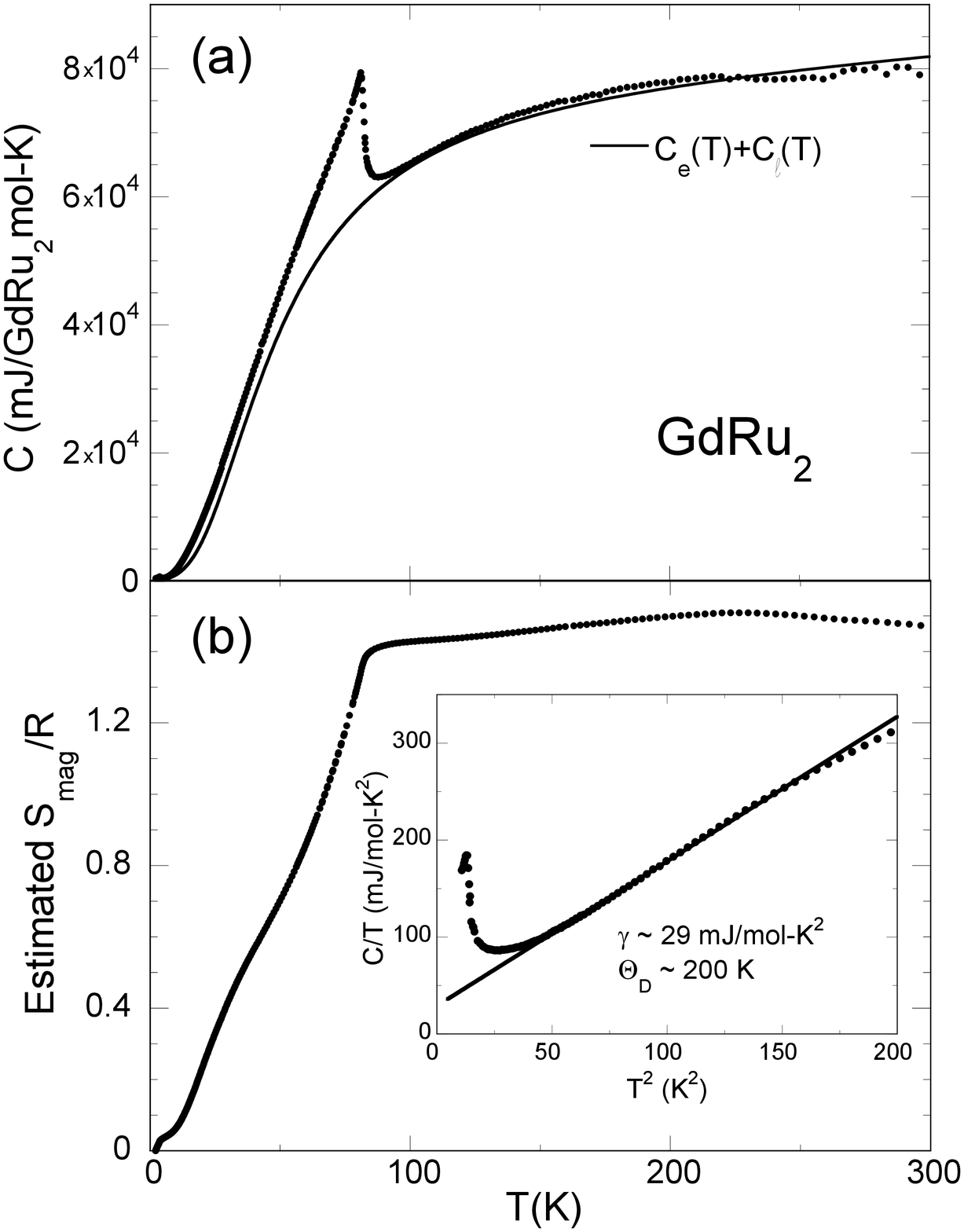}
 \caption{(a) Specific heat $C$ vs temperature $T$ of GdRu$_2$.
           The solid line is the sum of the estimated contributions
           from band electrons and phonons $C_{\rm{e}}(T)+C_{\rm{\ell}}(T)$.
           Inset of (b) $C/T$ vs $T^2$ below 14\,K.  The estimated values of
           electronic specific heat coefficient $\gamma$ and Debye temperature
           $\Theta_D$ are \mbox{$\sim 29$\,mJ/mol-$K^2$} and $\sim 200$\,K,
           respectively. (b) Estimated magnetic entropy $S_{\rm{mag}}$ of
           GdRu$_2$ after the contributions from band electron and phonons is
           removed.  The saturation value of $S_{\rm{mag}}$ is $\sim 1.5$R,
           less than Rln8.}
 \label{CvsTGdRu2}
 \end{center}
\end{figure}

Figure~\ref{CvsTGdRu2}(a) shows the specific heat $C$ of GdRu$_2$
from 2\,K to 300\,K. As $T$ decreases to $\sim 87$\,K, $C$ starts to
rise and peaks at $\sim 81$\,K, indicating the
ferromagnetic second order phase transition.
From the $C/T$ vs $T^2$ analysis between 7\,K and 14\,K  (inset of
Figure~\ref{CvsTGdRu2}(b)), the values of the electronic specific
coefficient $\gamma$ and the Debye temperature $\Theta_{\rm{D}}$ of
GdRu$_2$ are estimated to be 29\,mJ/mol-K, and 200\,K, which seem
reasonable in comparison with LaRu$_2$'s $\gamma \sim
41.6$\,mJ/mol-K and $\Theta_{\rm{D}} \sim
158.4$\,K.~\cite{Gschneider1972} After subtraction of electron and
phonon contributions $C_{\rm{e}}(T)+ C_{\rm{\ell}}(T)$ from C(T)
(Fig.~\ref{CvsTGdRu2}(a)), the temperature dependence of the
estimated magnetic entropy $S_{\rm{mag}}$ is displayed in
Fig~\ref{CvsTGdRu2}(b). Estimated $S_{\rm{mag}}$ reaches a saturated
value of 1.5\,R above T$_{\rm{C}}$ which is lower than the expected
value of Rln8 ($\approx$ 2R). This could be  due to a temperature dependent $\Theta_{\rm{D}}$
of GdRu$_2$ or our  overestimate for the
phonon contribution to the specific heat.

A HoRu$_2$ sample was synthesized so that magnetic order could also be investigated using neutron diffraction.
Gd has a large cross-section for neutron capture which makes investigations of GdRu$_2$ with neutron diffraction
impractical.
\begin{figure}
 \begin{center}
 \includegraphics[angle=270,width=1\textwidth]{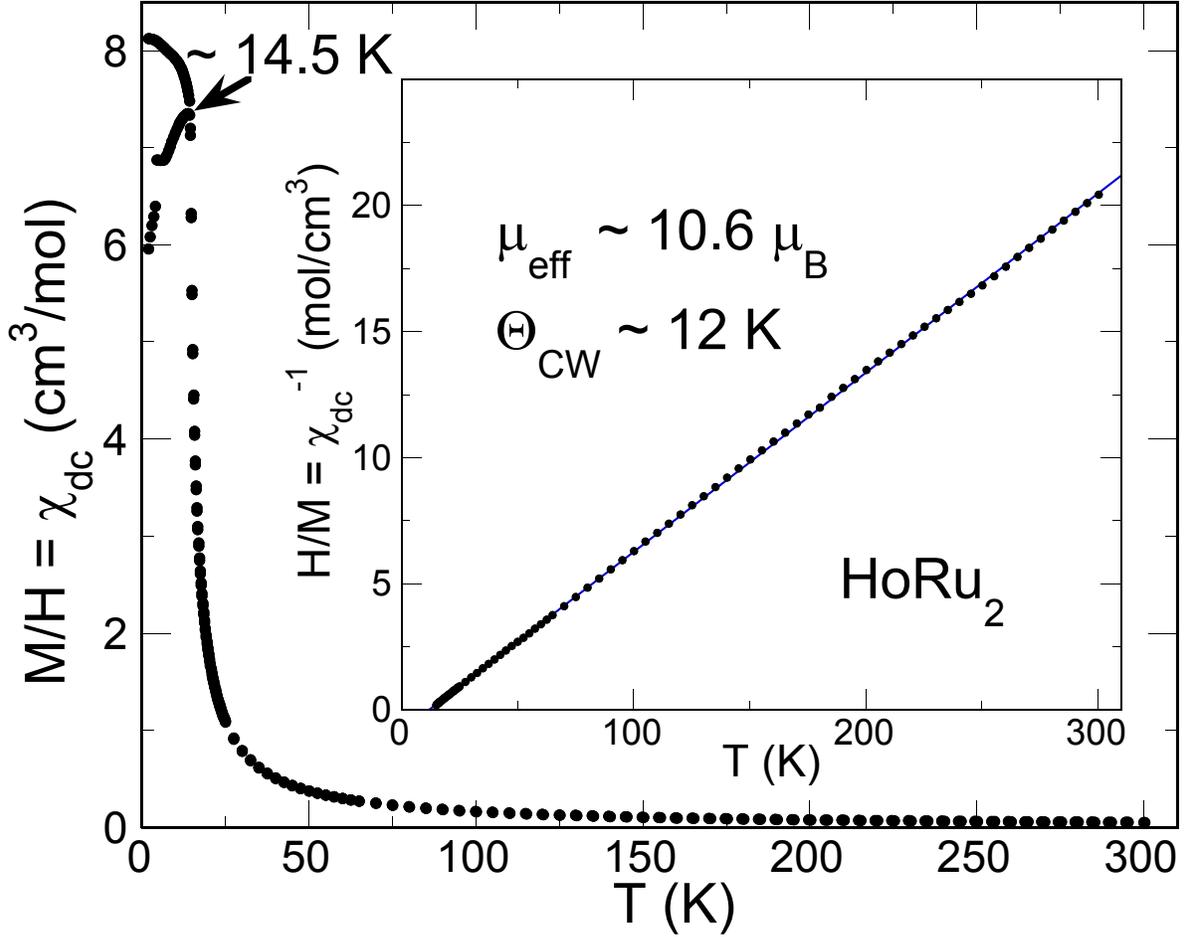}
 \caption{Temperature dependence of dc-magnetic susceptibility $\chi_{\rm{dc}}$ of HoRu$_2$
          measured at an applied magnetic field $H = 100$\,Oe.
          Hysteresis occurs at $T \sim 14.5$\,K.
          Inset $\chi_{\rm{dc}}^{-1}$ vs $T$ fits well with a Curie-Weiss behavior, which results in
          an effective moment $\mu_{\rm{eff}} \sim 10.6\mu_{\rm{B}}$ and
          a Curie-Weiss temperature $\Theta_{\rm{CW}} \sim 12$\,K.}
 \label{chidcvsTHoRu2}
 \end{center}
\end{figure}
Measurements of $\chi_{\rm{dc}}$, performed on HoRu$_2$ at $H$ =
100\,Oe and from 2 K to 300 K in the ZFC and FC conditions,
are displayed in Fig.~\ref{chidcvsTHoRu2}.
Hysteresis in $\chi_{\rm{dc}}(T)$ occurs at $\sim 14.5$\,K. The
Curie-Weiss analysis (shown in the inset of
Fig.~\ref{chidcvsTHoRu2}) indicates a ferromagnetic transition takes
place near $\Theta_{\rm{CW}} \sim 12$\,K in HoRu$_2$ with a
$\mu_{\rm{eff}} \approx 10.6$\,$\mu_B$ , which agrees with the
theoretical value 10.6\,$\mu_B$ of the free ion moment of Ho$^{3+}$.
An interesting feature in the ZFC $\chi_{\rm{dc}}$ data of HoRu$_2$
is that there is  a $20\%$ drop of the $\chi_{\rm{dc}}$ values at 2 K compared
to the value at 14.5\,K (similarly a $\sim 13\%$ drop in the ZFC
$\chi_{\rm{dc}}$ of GdRu$_2$, Fig.~\ref{chidcvsTGdRu2}).

\begin{figure}
\begin{center}
\includegraphics[width=.9\textwidth,height=0.8\textheight]
{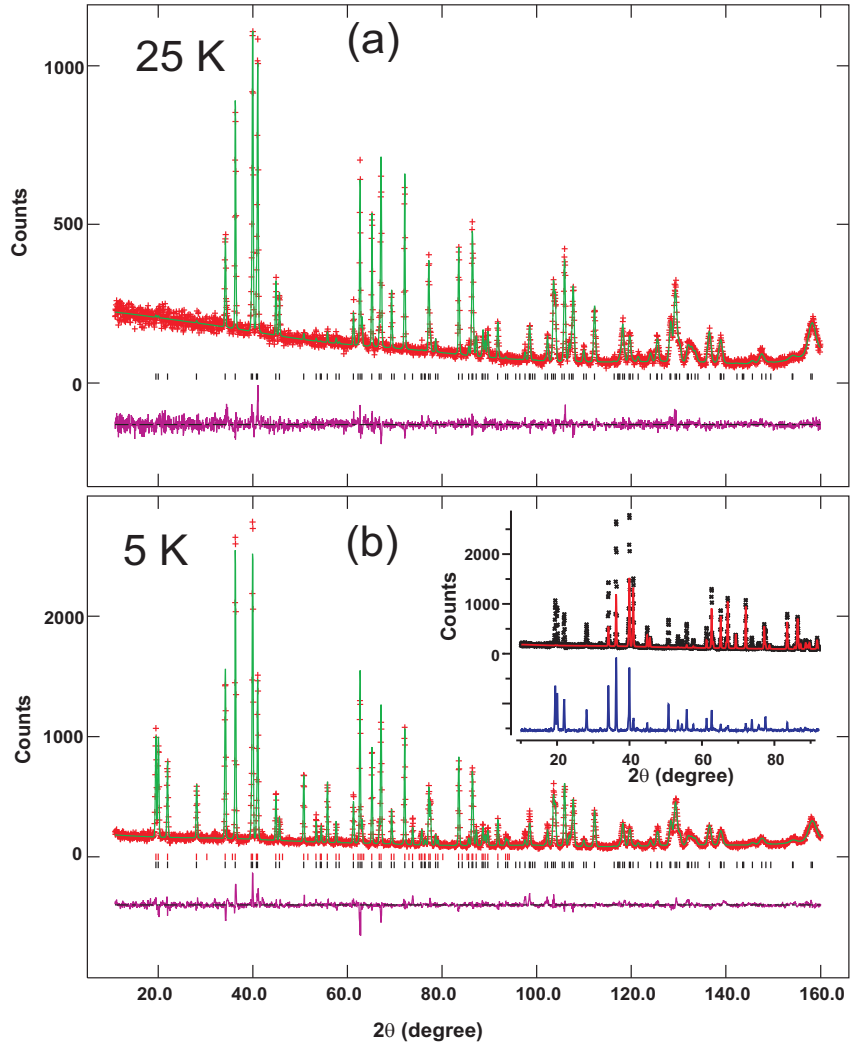}
\caption{(Color Online)High resolution neutron diffraction patterns taken (a) above and (b)
below the magnetic phase  transition at 15.3 K. The crosses indicate the observed data
and the solid curves the calculated intensities from the structural refinements,
and the difference is shown in the low part of the plots.
The vertical lines indicate the angular positions of the diffraction lines for the nuclear Bragg peaks, and (b) for the magnetic (top) Bragg peaks.}
\label{DiffractionPatterns}
\end{center}
\end{figure}
High resolution powder diffraction data were collected at the NCNR on the 
BT-1 high-resolution neutron powder diffractometer, using monochromatic neutrons 
of wavelength {1.5403{\AA}} produced by a Cu(311) monochromator.  S\"{o}ller collimations 
before and after the monochromator and after the sample were 15′, 20′, and 7′ 
full-width-at-half-maximum (FWHM), respectively.  
Data were collected in the $\theta$ range of 3$^o$ to 168$^o$ with a step size of 0.05$^o$ 
at 25K and 5K, above and below the magnetic phase transition.  
Structural refinements were carried out using the GSAS program.\cite{Larson1990}

\noindent
\begin{table}
\caption{\label {tab:table1} Refined crystal structure parameters for HoRu$_2$ at 25 K (first line) and
5 K(second line)
Space group {\it {P6/3mmc}}(No.194)
$a$=5.2310(2)(\AA), $c$=8.8265(4)(\AA), $V$=209.16(2)(\AA$^3$).
}
\vskip 4pt
\begin{tabular}{|c|c|c|c|c|c|c|c|}
\hline
 Atom    & x  & y  & z        & B(\AA$^2$) & M$_x$($\mu_B$) & M$_z$($\mu_B$)& M($\mu_B$) \\
\hline
 Ho      &1/3 & 2/3& 0.0664(2)&  0.36(3)&         &       &          \\
         &1/3 & 2/3& 0.0660(2)&  0.03(3)& 6.70(8) & 4.3(1)& 7.98(8)  \\
         &    &    &                    &         &       &       &        \\
 Ru$_1$  & 0  & 0  & 0        & 0.39(3) &         &       &          \\
         & 0  &  0 &  0       & 0.22(3) &         &      &          \\
         &    &    &                    &         &       &       &        \\
 Ru$_2$  & 0.1704(2)& 0.3407(2)& 3/4    & 0.39(3) &         &       &          \\
            & 0.1711(2)& 0.3422(4)& 3/4    & 0.22(3) &         &      &          \\
\hline
\end{tabular}
\vskip 4pt
\noindent
\hskip 15pt {\it {Rp}}=6.16\%,\hskip 45pt {\it {wRp}}=7.34\%,\hskip 45pt $\chi^2$=0.8387
\\ 
\noindent
\hskip 30pt 7.12    \hskip 80pt 9.02 \hskip 75pt 1.428
\end{table}

Detailed temperature dependent measurements of the magnetic order parameter were carried 
out on the BT9 triple axis spectrometer.  A pyrolytic graphite (PG) (002) 
monochromator was employed to provide neutrons of wavelength 2.36 \AA, 
and a PG filter was used to suppress higher-order wavelength contaminations.  
Coarse collimations of 40′, 48′, and 40′ FWHM on BT9 were employed to maximize the intensity.  
A PG(002) energy analyzer was used in these measurements.  
Inelastic measurements were taken with a fixed final energy of 14.7 meV.

\begin{figure}
\begin{center}
\includegraphics[width=.95\textwidth,height=0.7\textheight]
{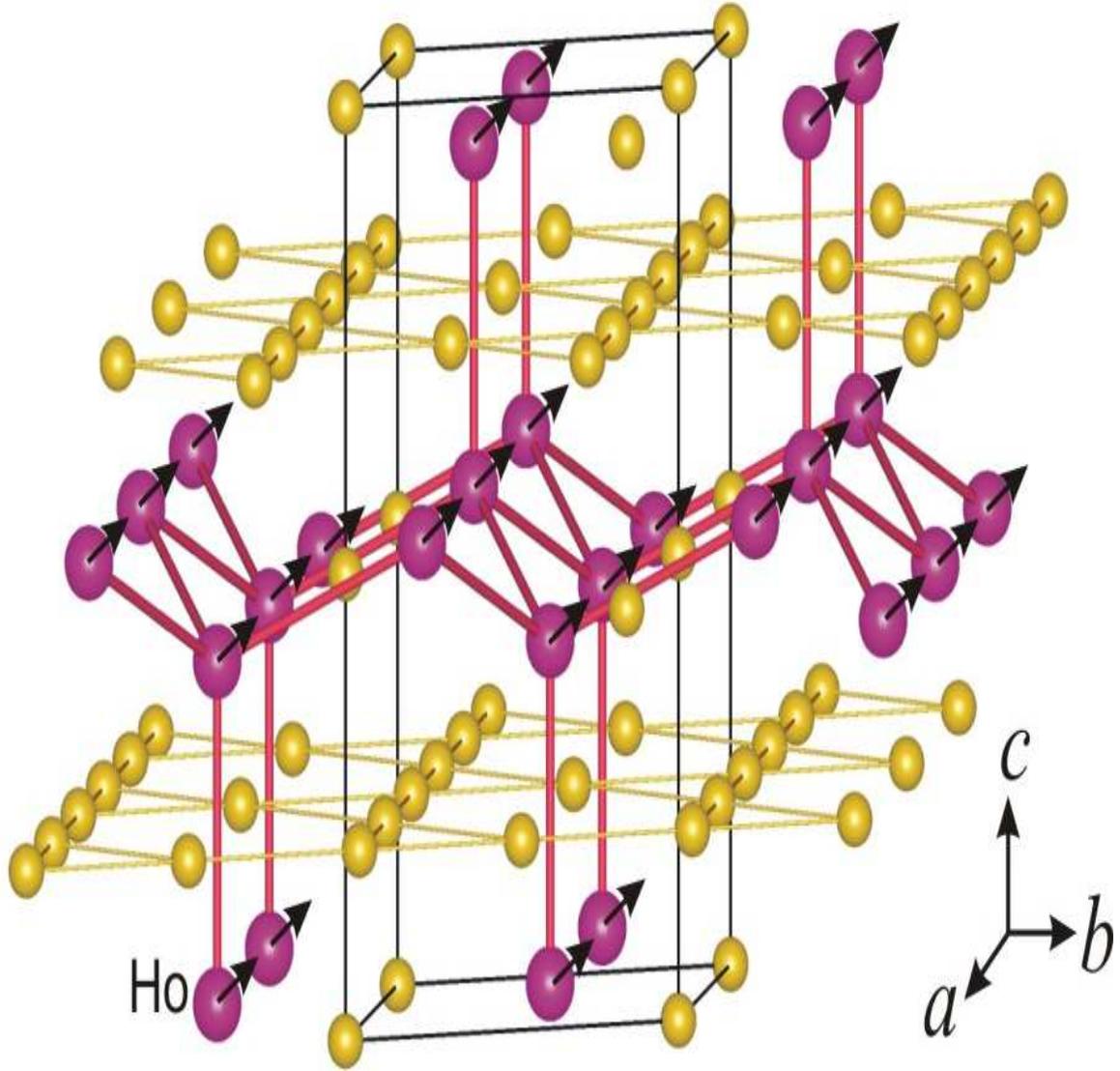}
\caption{(Color Online)The hexagonal laves phase and magnetic structure of HoRu$_2$ at 5 K.
The ferromagnetically ordered moments on the Ho sites have a magnitude of 7.98(8)$\mu_B$.
}
\label{NM_Cell_HoRu2}
\end{center}
\end{figure}

Fig. ~\ref{DiffractionPatterns}
shows the diffraction pattern obtained above (25K) and below (4 K) 
the magnetic transition.  The diffuse “background” scattering in the magnetically 
disordered state is due to paramagnetic scattering of the uncorrelated 
Ho moments, which decreases with increasing angle due to the magnetic form factor.  
The overall refinement fits at both temperatures are excellent.  
The structure is found to deviate slightly from the ideal phase, 
and the refined values for the crystal structure are given in Table 1.  
In the ground state the Ho ions exhibit long range ferromagnetic order, 
with both in-plane and c-axis components of the ordered moment also given in Table 1.  
We find an ordered moment of 7.98(8)$\mu_B$, 
which is in good agreement with the low temperature magnetization data of Andoh\cite{Andoh1987}
and somewhat smaller than the Curie-Weiss value obtained from 
the magnetization measurements. No evidence was found for moments on the Ru sites. 
The crystal and magnetic structures are shown in Fig. ~\ref{NM_Cell_HoRu2}.
\noindent
\noindent
\begin{figure}
\begin{center}
\includegraphics[width=.95\textwidth,height=0.7\textheight]
{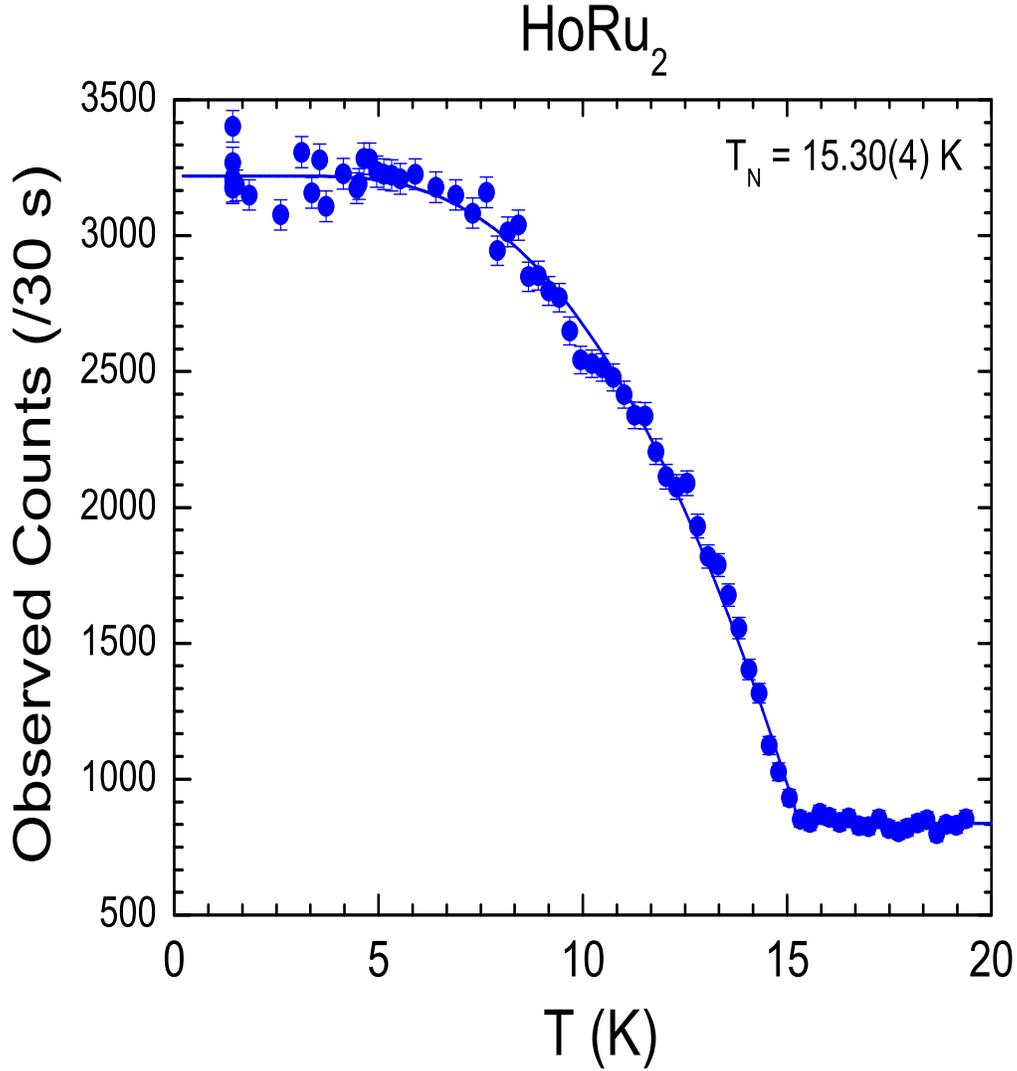}
\caption{(Color Online)Temperature dependence of the magnetic Bragg peak intensity,
which yields an ordering temperature of 15.30(4) K.  The intensity above the phase
transition originates from the nuclear Bragg peak.
Uncertainities are statistical in origin and represent one standard deviation.}
\label{BraggPeakHoRu2}
\end{center}
\end{figure}
\noindent
\begin{figure}
\begin{center}
\includegraphics[width=.95\textwidth,height=0.7\textheight]
{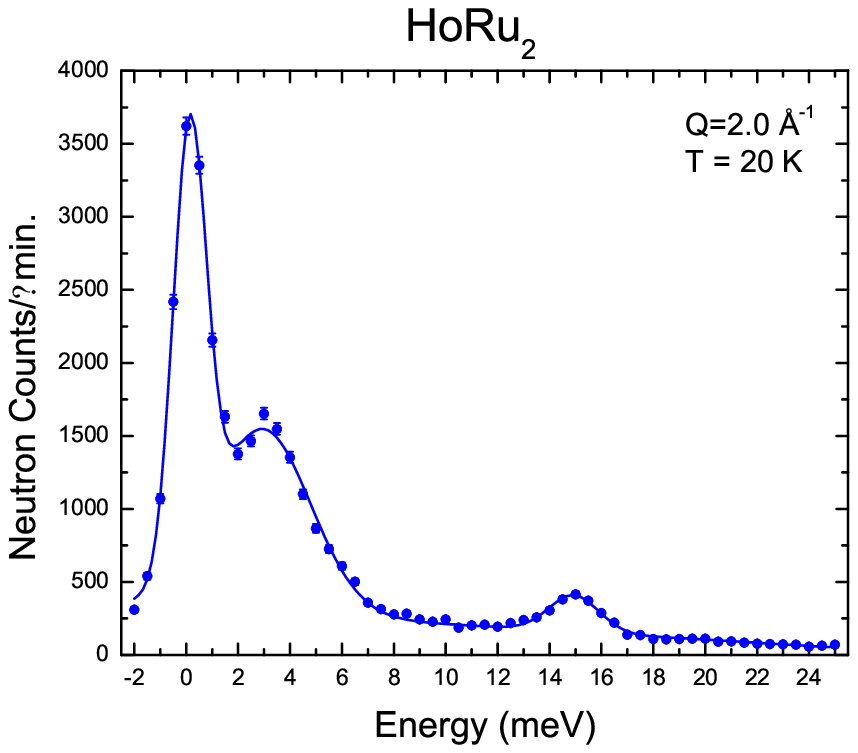}
\caption{(Color Online)Inelastic scattering observed at 20 K and
a wave vector of 2.0 \AA$^{-1}$.
The elastic peak has a magnetic component and nuclear incoherent scattering,
and is resolution limited. Two crystal field excitations are observed at
2.83(5) and 14.82(5) meV.
The low energy scattering is broad, with a width of 4.34(12) meV.
The energies and intensities are quite similar to the Ho crystal fields observed
in cubic Ce$_{0.73}$Ho$_{0.27}$Ru$_2$\cite{Lynn1979,Lynn1980}}.
\label{InelasticHoRu2}
\end{center}
\end{figure}

The temperature dependence of the magnetic Bragg intensity, 
which is proportional to the square of the ordered magnetic moment, is shown in Fig. 8. 
The temperature dependence is typical for magnetic ordering, 
and the solid curve 
is a least-squares fit of the intensity to a mean field order parameter, 
which provides a good fit to the data with a Curie temperature of 15.30(4) K.
\vskip 8pt
The diffraction data indicate that the ordered moment is reduced from the 
free-ion value of 10.0$\mu_B$, which suggests that crystal field effects 
are important in this system.  
We therefore carried out inelastic neutron 
scattering measurements on BT-7 for several temperatures and wave vectors to 
search for crystal field excitations.  Figure 9 shows a scan above the phase transition, 
at a temperature of 20 K.  We see two clear excitations from the crystal field ground 
state, a strong one at 2.83(5) meV and a weaker excitation at 14.82(5) meV.  
The energies and intensities turn out to be quite similar to the Ho crystal field 
levels observed in the closely related Ce$_{1-x}$Ho$_x$Ru$_2$ system for smaller 
$x$\cite{Lynn1979,Lynn1980},
which has the cubic C-15 Laves structure.  The rare earth site symmetry in 
the CeRu$_2$ case is cubic, $\bar{4}3m$, and the crystal field level 
scheme has been worked out in detail.  For hexagonal HoRu$_2$ the site symmetry is 
lower, $3m$, but the crystal field levels 
look remarkably similar nevertheless.  
The width for the higher energy level is limited by the 
instrumental resolution, while the level at 2.83 meV has an observed 
full-width-at-half-maximum (FWHM) of 4.34(12) meV, which is much broader than 
the resolution of 1.5 meV.  The width likely originates from exchange broadening.  
We note that both excitations are clearly magnetic in origin, 
as their intensity decreases with increasing wave vector, following 
the magnetic form factor dependence, and they decrease in intensity 
with increasing temperature as the ground state occupancy is depleted. 

All the thermodynamic, transport, neutron diffraction  measurements clearly
demonstrate that ferromagnetic order develops in GdRu$_2$ below 83\,K
and in HoRu$_2$ below 15\,K.

\subsection{$^{99}$Ru M\"{o}ssbauer Spectra }

Details of the source preparation of $^{99}$Rh(Ru), velocity calibration,  and the experimental setup
for transmission spectroscopy are discussed in previous 
papers\cite{DeMarco2000,Coffey2008}
The sample of CeRu$_2$ was made with natural Ru and contained 100mg/cm$^2$ of Ru.
Samples of GdRu$_2$, HoRu$_2$, and Ce$_x$Gd$_{1-x}$Ru$_2$ were prepared with enriched $^{99}$Ru(95\%).
Typical samples contained about 65mg/cm$^2$ of $^{99}$Ru.
This made it possible to measure
well-resolved spectra up to 150K.
In the analysis of the spectra the number of inequivalent Ru sites is different in these materials.
There is one type of Ru site in the unit cell of the cubic laves structure (CeRu$_2$),
whereas there are
two inequivalent Ru sites whose relative abundance is 1 to 3, 
in the hexagonal Laves structure (GdRu$_2$ and HoRu$_2$).
\noindent
\begin{figure}
\begin{center}
\includegraphics[width=.8\textwidth,height=0.35\textheight]
{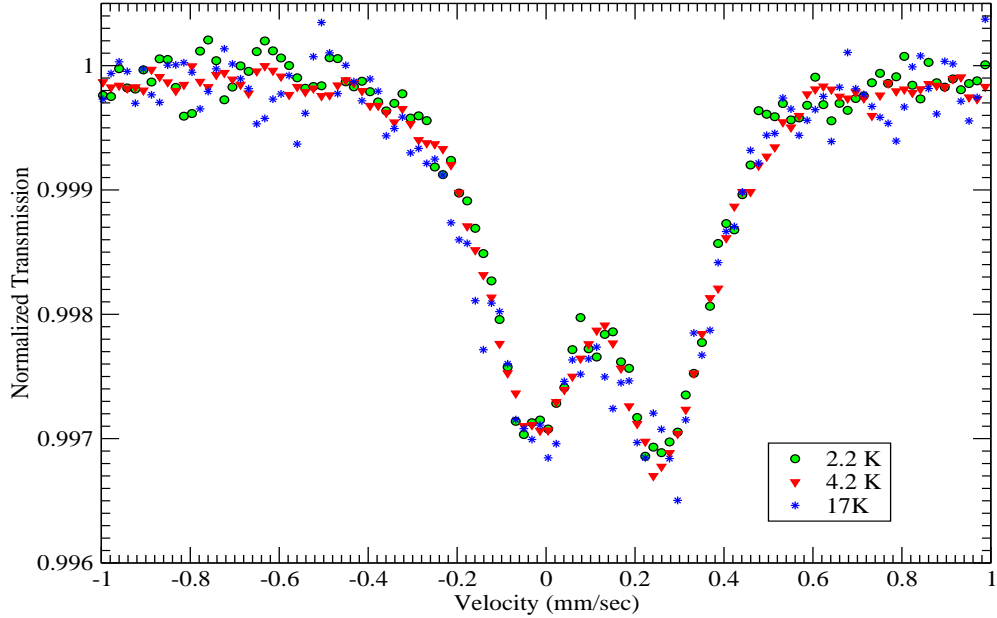}
\caption{(Color Online)
$^{99}$Ru M\"{o}ssbauer spectra of Ce$_{0.88}$Gd$_{0.12}$Ru$_2$ at 2.2 K(o),
4.2 K($\bigtriangledown$), and 17 K($\ast$).
Both the superconducting transition temperature and 
Curie temperature are $\sim$ 4 K\cite{Wilhelm1971}.
The temperature independent spectra show no evidence of a hyperfine magnetic field.}
\label{Ce88Gd12Ru-Temp}
\end{center}
\end{figure}

\begin{figure}
\begin{center}
\includegraphics[width=.8\textwidth,height=0.7\textheight,angle=-90]
{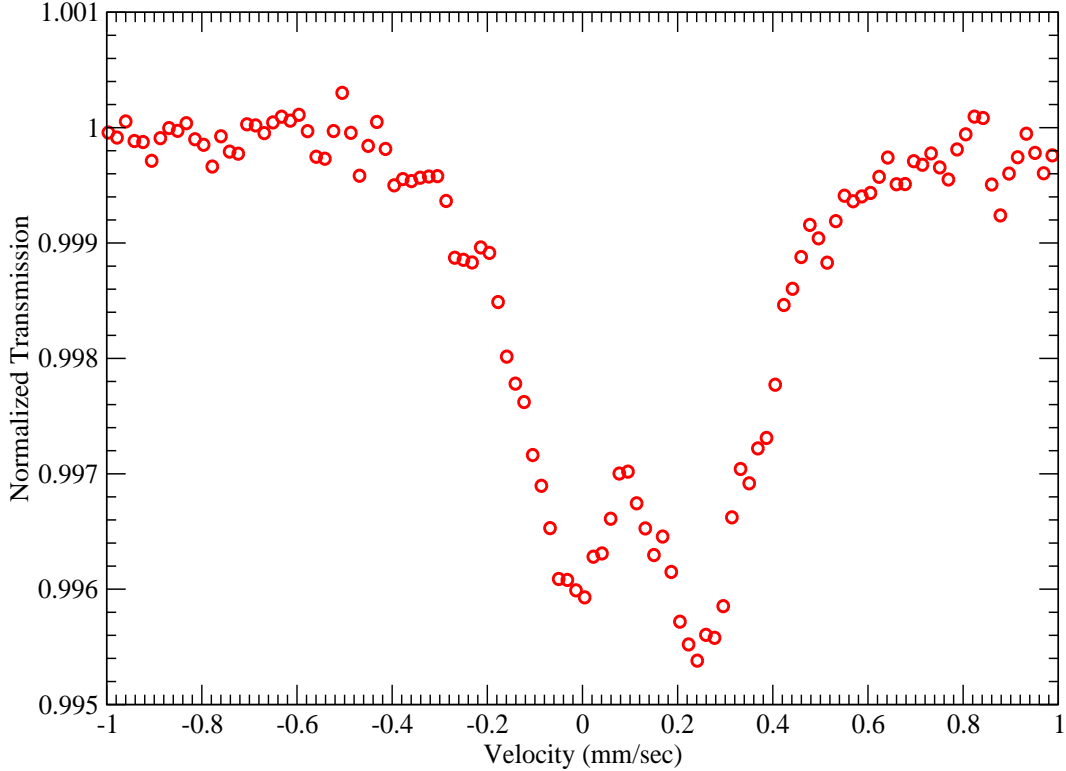}
\caption{(Color Online)$^{99}$Ru M\"{o}ssbauer spectrum of Ce$_{0.2}$Gd$_{0.8}$Ru$_2$ at 4.2 K.
Extrapolating the data of Wilhelm and Hillenbrand\cite{Wilhelm1971},
 T$_C$ for this material is $\sim$60 K.
This spectrum is very similar to that of Ce$_{0.88}$Gd$_{0.12}$Ru$_2$ except for additional asymmetry due to the increased Gd
concentration.
 There is no evidence for the hyperfine magnetic field at the Ru site expected in a ferromagnetically ordered material.
}
\label{Ce20Gd80Ru-4.2K}
\end{center}
\end{figure}

\noindent
\begin{figure}
\begin{center}
\includegraphics[width=.8\textwidth,height=0.7\textheight,angle=-90]
{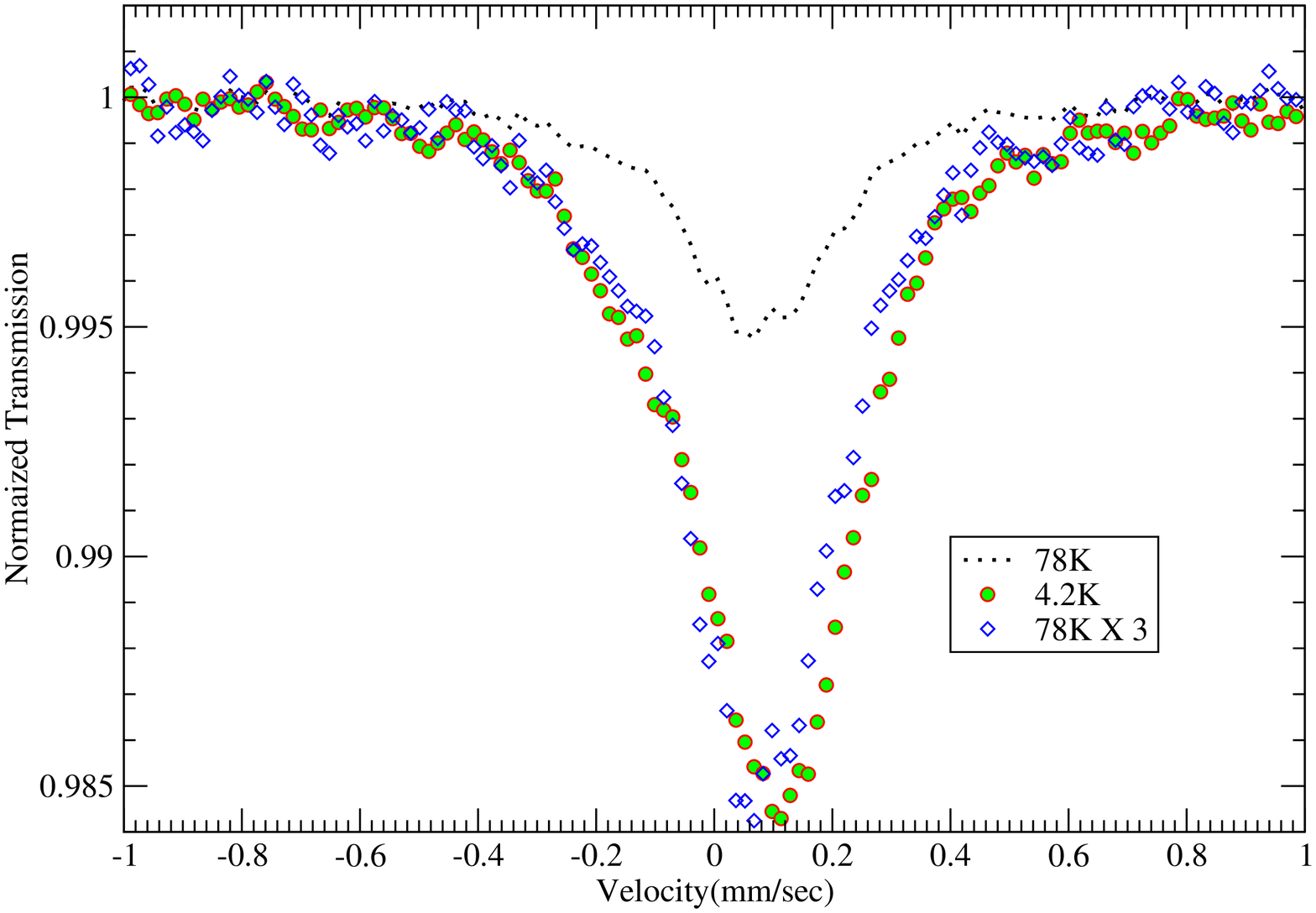}
\caption{(Color Online)$^{99}$Ru M\"{o}ssbauer spectrum of GdRu$_2$
        at 4.2 K (o) and at 78 K ($\diamond$)
for a sample whose Curie temperature is 88.6 K. The 78 K spectrum(line) is scaled by 3.0 for direct comparison with
the spectrum at 4.2 K.
The only temperature dependence in the spectrum between 4.2 K and 100 K,
based also on measured spectra at 89 K and 101 K, is a continuous shift of the isomer shift to
slightly more positive values. Again there is no evidence of a hyperfine magnetic field.}
\label{GdRu-Temp}
\end{center}
\end{figure}

CeRu$_2$ is a superconductor with a transition temperature, T$_{SC}$, of 6K.  
As the Gd content of Ce$_x$Gd$_{1-x}$Ru$_2$ increases with $x$,  
T$_{SC}$ falls and goes to zero at $x\simeq 0.14$. 
Previous authors have suggested that superconductivity and ferromagnetism could coexist close to
this doping level. The spectrum of Ce$_{0.88}$Gd$_{0.12}$Ru$_2$  at 2.2 K, 4.2 K and 17K is 
shown in Fig.~\ref{Ce88Gd12Ru-Temp}. 
Extrapolating the data of Wilhelm and Hillebrand\cite{Wilhelm1971}
gives values for T$_{SC}$ and T$_C$ both $\sim$4 K.
There is no evidence in the temperature independent spectra for a hyperfine magnetic field,
suggesting that ferromagnetism is ruled out at this doping.
However, as doping with Gd increases, 
the low temperature phase is known to be ferromagnetically ordered
and 
at $x=0.2$  one would expect to see an
eighteen line magnetic spectrum at 4.2 K due to a  large value of $B_{hyperfine}$
field at the Ru site.  Instead Fig.~\ref{Ce20Gd80Ru-4.2K} shows a quadrupole spectrum 
which is somewhat more asymmetric than at $x$=0.88  due to the increased 
Gd content. The absence of hyperfine magnetic field even in this sample suggests that the
spectra of Ce$_{0.88}$Gd$_{0.12}$Ru$_2$
do not rule out the coexistence of magnetism and superconductivity. 

The M\"{o}ssbauer spectrum for GdRu$_2$ is an almost temperature independent
single peak between 4.2 K and 101 K except for a shift toward more positive velocities.
This is shown in Fig.~\ref{GdRu-Temp} for 4.2 K and 78 K, where
the 78K spectrum is scaled by 3 to compensate for the temperature
dependence of the recoil free fraction so that a direct comparison can be made with the 
4.2 K spectrum. The experimental FWHM of the spectrum 
is slightly broader than that of Ru powder (0.25 mm/s).
The width of the spectrum is unchanged with temperature.  
This is also the case for the spectra at 89 K and 101 K (not shown).
Contrary to the transport, magnetic, and thermodynamic data on the same sample presented
in the previous section,
 there is no evidence of a transferred hyperfine field at the Ru sites due to magnetic order 
on the Gd sites.

The same apparent discrepancy between 
the M\"{o}ssbauer spectrum and the experiments discussed above is seen in 
HoRu$_2$.
The spectrum of HoRu$_2$ at 4.2 K is similar to that of GdRu$_2$. It is  a single peak
with no evidence of splitting due to a hyperfine magnetic field
even though the temperature is far below 
the sample's Curie temperature, 15.3K.
In order to get a more detailed picture of the properties of these materials we examined their 
electronic structure using {\it { ab initio}} calculations. \vskip 16pt
\section{Electronic Structure}

The lattice parameters and structure used in the calculations are those given
by Compton and Matthias\cite{Compton1959} for CeRu$_2$ and by the results of the neutron diffraction
measurements of HoRu$_2$ at 5 K discussed.
The lattice parameters and structure were measured by Compton and Matthias\cite{Compton1959}
for GdRu$_2$ at room temperature. However, the lattice frequently expands below a ferromagnetic transition 
in rare earth intermetallic compounds\cite{Ohta1995}.
So we determined lattice constants for GdRu$_2$ at low temperatures by scaling the lattice constants given by Compton and Matthias
and found that the total energy was minimised when their values were increased by $\sim 1$\%.

\subsection{Density of States and Magnetic Order}

The absence of the expected large splitting in the M\"{o}ssbauer spectra in GdRu$_2$ and HoRu$_2$ points 
to a very small values of $B_{hyperfine}$ given that the materials are ferromagnetically ordered.
This is completely different from the ruthenates where the dependence of the M\"{o}ssbauer spectra 
reflect the internal ordered field.\cite{DeMarco2000,Coffey2008}
In particular, SrRuO$_3$ orders ferromagnetically at 163 K and the value of $B_{hyperfine}$ is 
found to be 33 T at 4.2 K.

We have calculated the electronic structure 
and the values of $B_{hyperfine}$  at  the Ru and 
lanthanide sites using the Wien2k software package\cite{wien2k}.
These calculations use the LSDA, which is 
implemented using an extension of the augmented plane wave(APW)
method.
Up to 729 inequivalent $\vec k$ points were used in the P6$_3$/mmc structure and  RKMAX was set at 8
to ensure convergence.
The energy and charge convergence variables were $ec=10^{-5}$ Rydbergs and $cc=10^{-5}e$.
In each region a wavefunction basis set is chosen to optimize the calculation.  
The spin on each site, S, is the net electronic spin polarization in the 
sphere surrounding that site.  

The calculated contributions to the densities of states
in CeRu$_2$, GdRu$_2$, and HoRu$_2$
show that the $f$-bands are  narrow in each case, pointing to predominantly
localized states.
In CeRu$_2$ the calculated ground state is paramagnetic and the f-band is mostly above
the Fermi energy, with small weight at E$_F$ so that the {\it f} electrons
also contribute to the conduction band.
This is in agreement with the earlier {\it ab initio} calculations
of Yanase\cite{Yanase1986}
and Higuchi and Hasegawa\cite{Higuchi1996} and with experiment.

\begin{figure}
\begin{center}
\includegraphics[width=.9\textwidth,height=0.8\textheight]
{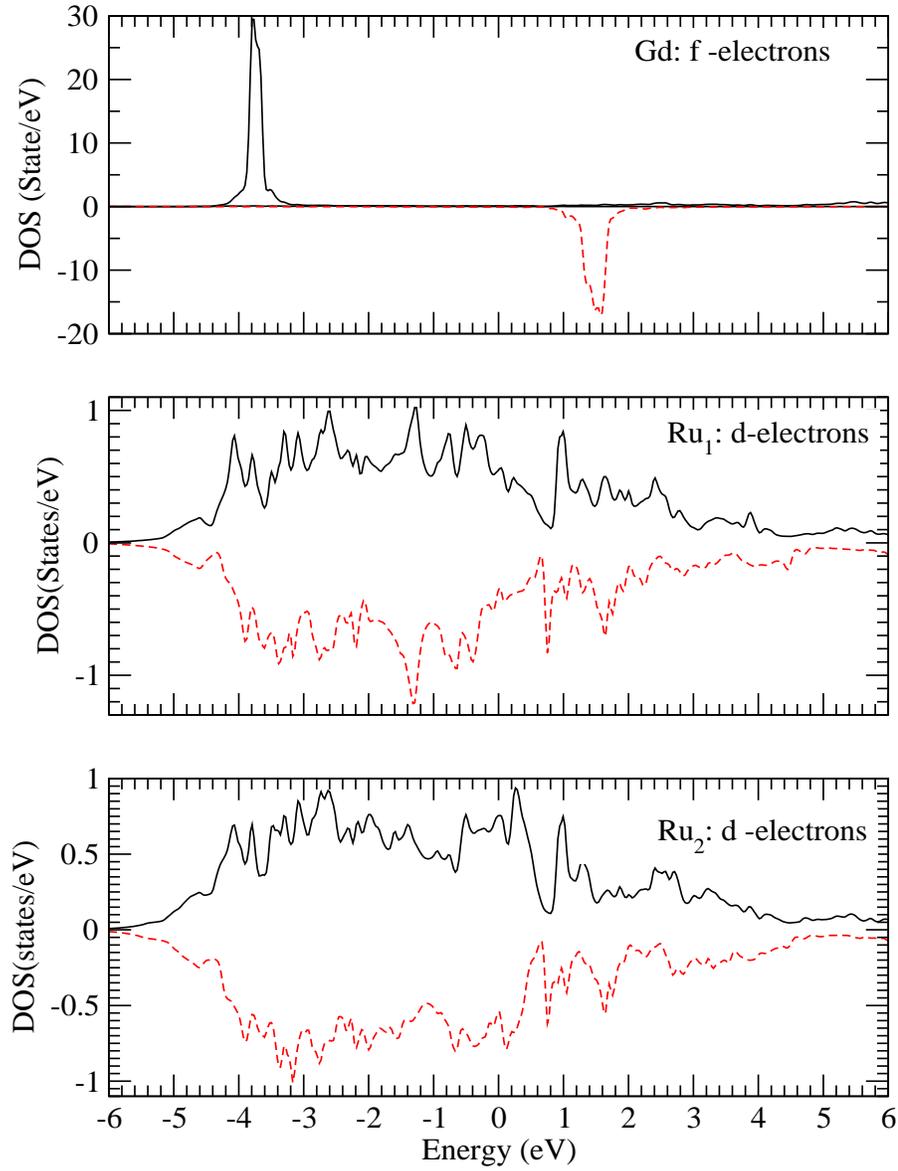}
\caption{(Color Online) Contributions to the Density of States in GdRu$_2$.
Lines are spin up and dashed-lines are spin down. E=0 is the Fermi level.}
\label{Dos1}
\end{center}
\end{figure}

\begin{figure}
\begin{center}
\includegraphics[width=.75\textwidth,height=0.7\textheight,angle=-90]
{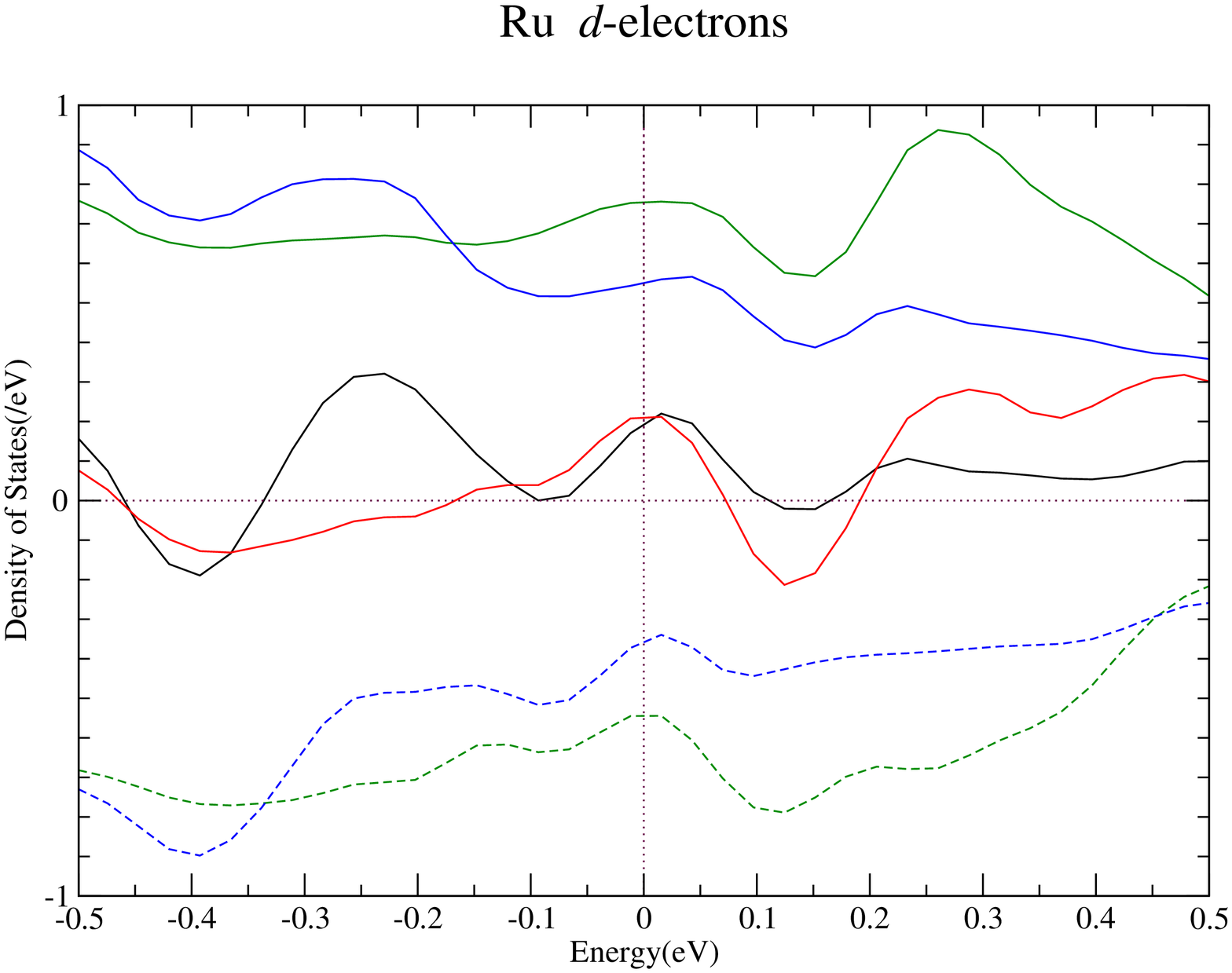}
\caption{(Color Online) Polarization of $d$-electrons from the two inequivalent Ru sites.
Black line - $N_{\uparrow}-N_{\downarrow}$ from Ru site-1.
Red line - $N_{\uparrow}-N_{\downarrow}$ from Ru site-2.
Blue and green lines above zero line are spin up and
the corresponding dashed-lines below the zero line are spin down. E=0 is the Fermi level.}
\label{Dos2}
\end{center}
\end{figure}

The calculated contributions to the density of states of  GdRu$_2$ 
from the Gd 4$f$ band and from the 4$d$ bands  associated with the two
inequivalent Ru sites sites are shown in Fig.~\ref{Dos1}.
The Gd spin up {\it f}-band is $\sim$4 eV below the Fermi level
and spin down {\it f}-band is  $\sim$1.5 eV above. 
The calculated value of net spin, S,  on the Gd sites is
3.46 which is 
very close to the Hund's rule result for Gd, $\frac{7}{2}$,
 and to the result 
of the Curie-Weiss fit to the magnetic susceptibility.
The calculated values of S at the two inequivalent Ru sites in GdRu$_2$ are  very small,
$\sim$-0.06 and $\sim$-0.09.
The bottom two panels in Fig.~\ref{Dos1} show the contributions to the density of states from Ru 4$d$ electrons
at the two inequivalent Ru sites.
$N_{\uparrow}$ is the contribution for states with spin parallel to the
moment of the Gd site and  is positive, and the contribution from spin down, $N_{\downarrow}$,
 is negative for each of the inequivalent Ru two sites.
 At the Fermi energy the density of states is dominated by the Ru 4$d$ electrons, 
which form broad conduction bands. 
The contributions from Ru 4$d$-electrons, $N_{\uparrow}$ and $N_{\downarrow}$, and  
$N_{\uparrow}-N_{\downarrow}$ are plotted in Fig.~\ref{Dos2} in the energy range
which is $\sim$0.1 eV on either side of the Fermi energy, E$_F$=0.
It is seen that
the $d$-electrons from both Ru sites 
are polarized parallel to the Gd moment with 
$\frac{N_{\uparrow}-N_{\downarrow}}{N_{\uparrow}+N_{\downarrow}} \simeq$0.2 
for each band arising from the
two inequivalent Ru sites. The total Ru $d$ density of states $\sim$1.0 (eVRu)$^{-1}$ at E=E$_F$. 

On the other hand the Gd $s$-electrons
have negligible weight at E$_F$, $\simeq$0.02 (eVGd)$^{-1}$ and are unpolarized.
This is also the case for Ru $s$-electrons.
This suggests that Ru $d$-electrons mediate the interaction between localized Gd $f$-moments.
In HoRu$_2$, the Ru 4$d$ Ru electrons play the same role but the minority
spin band is at the Fermi energy so that it is partially occupied.

In HoRu$_2$,  the
value of S at the Ho site is $\simeq $1.82.
This value of S is slightly lower than 2,
the value consistent with
the effective moment given by the fit to magnetic susceptibility,
as determined by Hund's Rules.
The absence of moments on the Ru sites in HoRu$_2$, S$\simeq$0, is consistent with
the neutron diffraction data.

According to these results, the magnetic properties 
can be described by a model 
in which localized Gd or Ho $f$-moments couple to 
itinerant Ru $d$-electrons, rather than to itinerant $s$ electrons\cite{Dunlap1973}.
\noindent
\noindent
\noindent
\begin{figure}
\begin{center}
\includegraphics[width=.8\textwidth,height=0.45\textheight]
{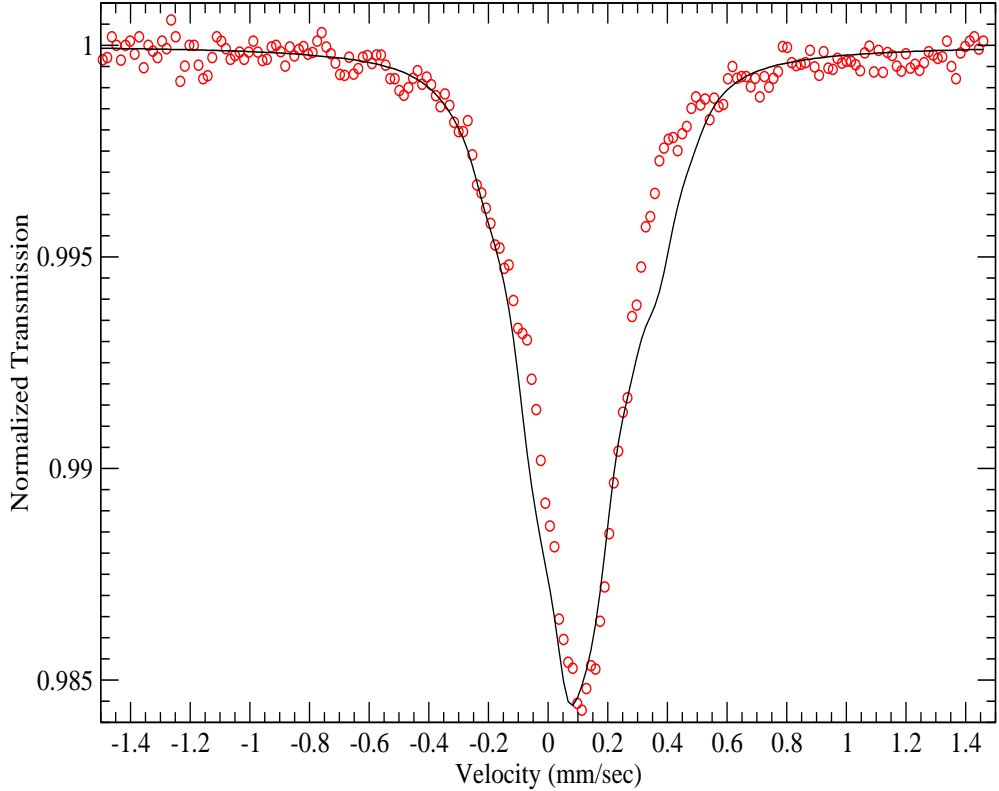}
\vskip 8pt
\caption{(Color Online)M\"{o}ssbauer spectrum of GdRu$_2$ at 4.2 K (o)
for a sample whose Curie temperature is 83.2 K. The solid line is the spectrum calculated with
the electric field gradient components and the B$_{hyperfine}$, shown in Table 2 and 4.
The
fitting parameters are the isomer shifts at the two inequivalent Ru sites, IS$_1$=0.0 mm/sec and IS$_2$=0.12 mm/sec, and the half-width of the lines, $\Gamma$=0.07 mm/sec.}
\label{GdRu-4.2K}
\end{center}
\end{figure}
\noindent
\begin{figure}
\begin{center}
\includegraphics[width=.8\textwidth,height=0.45\textheight]
{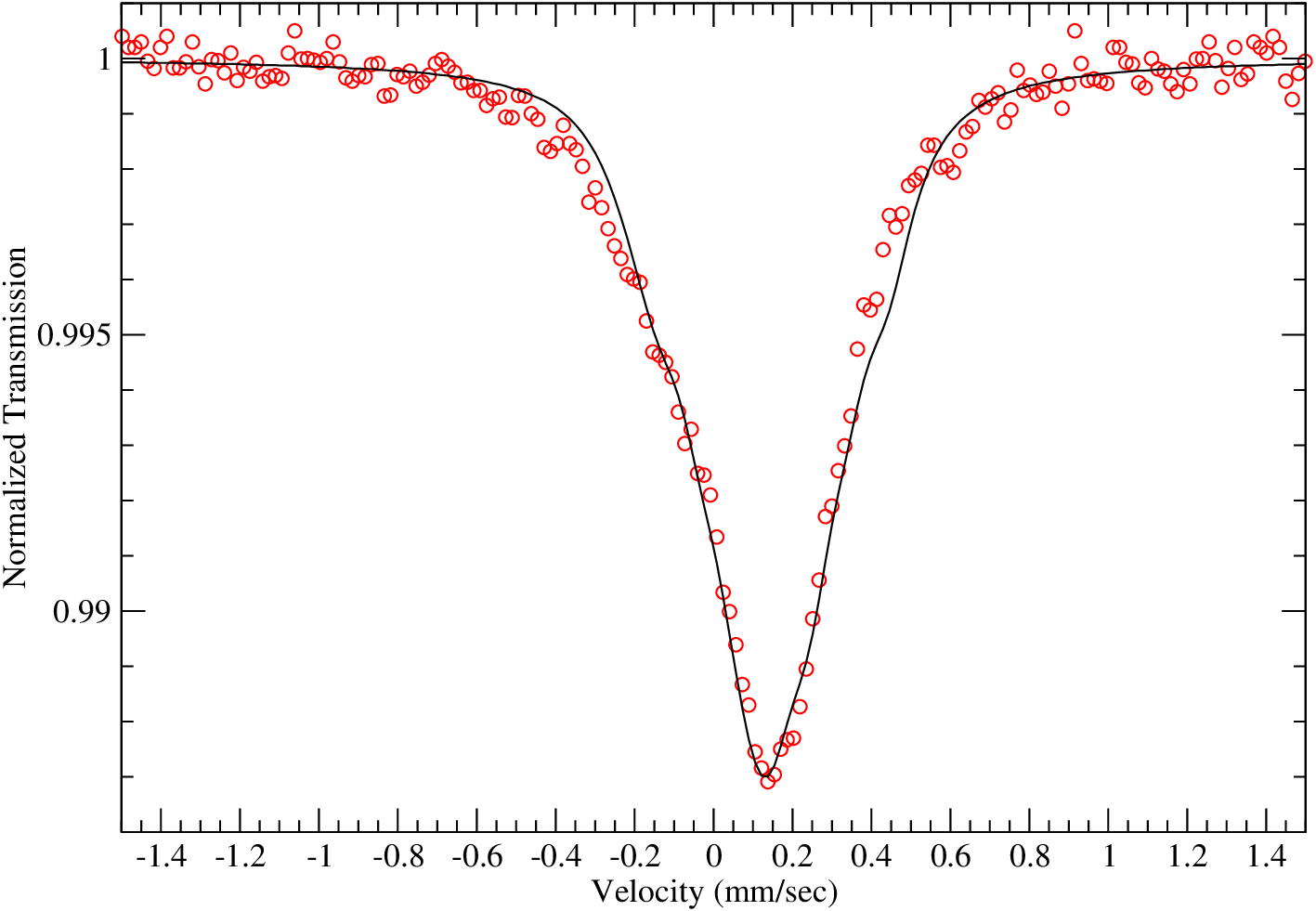}
\vskip 8pt
\caption{(Color online) $^{99}$Ru M\"{o}ssbauer spectrum of HoRu$_2$ at 4.2 K.
The Curie temperature is 15.3 K for this sample.
The solid line is the spectrum calculated with
the electric field gradient components and the B$_{hyperfine}$, shown in Table 2 and 4.
The
fitting parameters are the isomer shifts at the two inequivalent Ru sites,
IS$_1$=0.05 mm/sec and IS$_2$=0.18 mm/sec, and the half-width of the lines, $\Gamma$=0.08 mm/sec.}
\label{HoRu-4.2K}
\end{center}
\end{figure}
\noindent
\subsection{Hyperfine Magnetic Fields and Electric Field Gradients}

B$_{hyperfine}$ at a given site has a number of contributions which have been derived
with relativistic corrections by
Bl\"{u}gel et al.\cite{Blugel1987}
 and can be calculated directly using the Wien2k package\cite{Novak2006}.
These contributions are B$_{con}$=B$_{core}$+B$_{valence}$, the Fermi contact term,
B$_{dip}$, the dipolar field from the
on-site spin density, B$_{orb}$, the field associated with the on-site orbital moment,
and B$_{lat}$ , the classical dipolar field from all other atoms
in the system carrying moments.
B$_{core}$ is the contribution due to polarization of core electrons and
B$_{valence}$ is the contribution from the polarization of the valence or conduction band electrons.
These are by far the largest contributions.
Powder samples are used in the experiments with small crystallites
so that each Ru nucleus sees the average of a collection of randomly orientated
dipolar fields which presumably sum to zero.
Therefore, we take B$_{lat}$ to be zero in the current analysis.

The calculated B$_{hyperfine}$  and local moments at the Ce and Ru sites in CeRu$_2$ 
are essentially zero,
consistent with the absence of magnetic order.
The calculated contributions to B$_{hyperfine}$ are shown in Table 2 
for GdRu$_2$ and in Table 3 for HoRu$_2$.
The values of B$_{hyperfine}$ at the Ru sites are surprisingly small 
compared to that at the rare earth sites in both compounds, 
consistent with the M\"{o}ssbauer data.
By comparison, the transferred hyperfine field at the Ir site in Ir$_{0.01}$Fe$_{0.99}$ is 14.3 T
compared to 32 T at the Fe site\cite{Atzmony1967}.
The different 
 contributions to B$_{hyperfine}$ on the Ru sites are all small, leading to modest values
for B$_{hyperfine}$.  
The values of B$_{hyperfine}$  at the Ru nuclei are sensitive to the 
value of the lattice constants. 
In order to determine the lattice constants for GdRu$_2$ at low temperature, 
the electronic properties were calculated for values of the lattice constants given
by Compton and Matthias\cite{Compton1959}
scaled by 
a common factor, $\alpha$, from 0.995 to 1.05.
Although the value of the net spin on the Gd site did not change for this range of lattice constants,
the hyperfine magnetic fields at the Ru sites varied from -4.49 T at $\alpha=1.00$ to
0.1 T at $\alpha=1.05$.
The lowest total energy for a formula unit occurs at $\alpha=1.01$ and the
values for GdRu$_2$ are calculated with this $\alpha$. 
From Table 2, B$_{hyperfine}$ at both Ru sites is negative,
as one would expect from transferred hyperfine fields\cite{Watson1961a,Watson1961b}.
\noindent
\begin{table}
\caption{\label{tab:table2}
GdRu$_2$:  Net electron spin, S,  and contributions to the hyperfine magnetic fields(T) at the Gd and Ru sites
}
\begin{ruledtabular}
\begin{tabular}{|c|c|c|c|c|c|c|}
         & S   & B$_{core}$ & B$_{valence}$ & B$_{orb}$ & B$_{dip}$ & B$_{hyperfine}$ \\
\hline
 Gd      & 3.46   & 334.55    & -285.51       &  -2.88  & -.29     & 49.05         \\
 Ru$_1$  & -0.06 & -4.83    & 0.63         & -.08    & 0.01     & -4.20         \\
 Ru$_2$  & -0.09 & -3.78     & 1.06         & -.0.20    & 0.05     & -2.72         \\
\end{tabular}
\end{ruledtabular}
\end{table}
\noindent
\begin{table}
\caption{\label{tab:table3}
HoRu$_2$:  Net electron spin, S,  and contributions to the hyperfine magnetic fields(T) at the Ho and Ru sites
}
\begin{ruledtabular}
\begin{tabular}{|c|c|c|c|c|c|c|}
         & S   & B$_{core}$ & B$_{valence}$ & B$_{orb}$ & B$_{dip}$ & B$_{hyperfine}$ \\
\hline
 Ho      & 1.82 & 179.32   & -161.31     & 38.72   &-.41      &56.34         \\
 Ru$_1$  & 0.028 &-1.63      & -0.26         & -0.02   & 0.01     & -1.90        \\
 Ru$_2$  & -0.009 &-2.61      & 0.69         & -0.75    & -0.070    & -2.74        \\
\end{tabular} 
\end{ruledtabular}
\end{table}
\noindent
\begin{table}
\caption{\label{tab:table4}
Diagonal elements of the Electric field gradient tensor (10$^{21}$V/m$^2$) at different sites
in GdRu$_2$  and HoRu$_2$.
}
\begin{ruledtabular}
\begin{tabular}{|c|c|c|c|c|c|c|c|}
\hline
  Ce  & (0.,0.,0.)        & & Gd    & (-1.40,-1.40,2.79) & & Ho     &(-1.34,-1.34,2.68) \\
  Ru  & (2.80,2.80,-5.60) & & Ru$_1$ & (3.42,3.42,-6.83) & & Ru$_1$ &(3.85,3.85,-7.71) \\
      &                   & & Ru$_2$ & (1.86,0.90,-2.76) & & Ru$_2$ &(1.30,0.90,-2.19) \\
\end{tabular}
\end{ruledtabular}
\end{table}

Examining the contributions 
to B$_{hyperfine}$ on the rare earth  site 
one sees that $B_{core}$ and $B_{valence}$ are large and of opposite sign.
This  is again consistent with the original discussion of Watson and Freeman\cite{Watson1961a,Watson1961b}.
The other contributions are negligible by comparison in GdRu$_2$.
However, in the HoRu$_2$ site $B_{orb}\sim$40 T,
 is a substantial fraction of B$_{hyperfine}$
on the Ho site due to the almost complete cancellation of the $B_{core}$ and $B_{valence}$. 
 This difference in the value of $B_{orb}$  at the Gd and Ho sites
is a reflection of the difference in 
the orbital angular momentum quantum number, $L$, on the Gd and Ho sites.  
Whereas Hunds' rules give $L=0$ on the Gd site because of the half-filled 4$f$ shell,
they give $L=6$ on the Ho site. 

The calculated components of the EFG tensor for CeRu$_2$, GdRu$_2$, and HoRu$_2$ are given in Table 3.
Since the Ce site symmetry is cubic the electric field gradient is zero.
The components of the EFG at the Ru site, V$_{xx}$=V$_{yy}$=2.80$\times$10$^{21}$Vm$^{-2}$,
are small compared to those in RuO$_2$\cite{Kistner1966}, which also has a pure quadrupole spectrum,
suggesting that the Ru sites have almost cubic symmetry.
In GdRu$_2$ and HoRu$_2$, the EFG are different at the two inequivalent Ru sites.
Again the magnitude of the components of the EFG suggests that there is almost cubic symmetry at these sites.

The M\"{o}ssbauer spectra for these materials can now be almost completely
determined with these calculated 
hyperfine magnetic fields and electric field gradients.
The additional parameters are the half-width half-maximum, $\Gamma$, of the absorption lines and the
isomer shifts at the Ru sites.
These are choosen to fit the data.
The GdRu$_2$ spectrum at 4.2 K is shown in Figure \ref{GdRu-4.2K}
and that of HoRu$_2$ in Figure \ref{HoRu-4.2K}.
The line through the spectrum is determined by
the EFG tensor and values of $B_{hyperfine}$ at the two inequivalent Ru sites
calculated with the parameters in Table 1, Table 2, and Table 3.
The non-Lorenzian shape is due to the calculated
difference in the EFG's at the two sites.

\begin{figure}
\begin{center}
\includegraphics[width=.9\textwidth,height=0.7\textheight,angle=-90]
{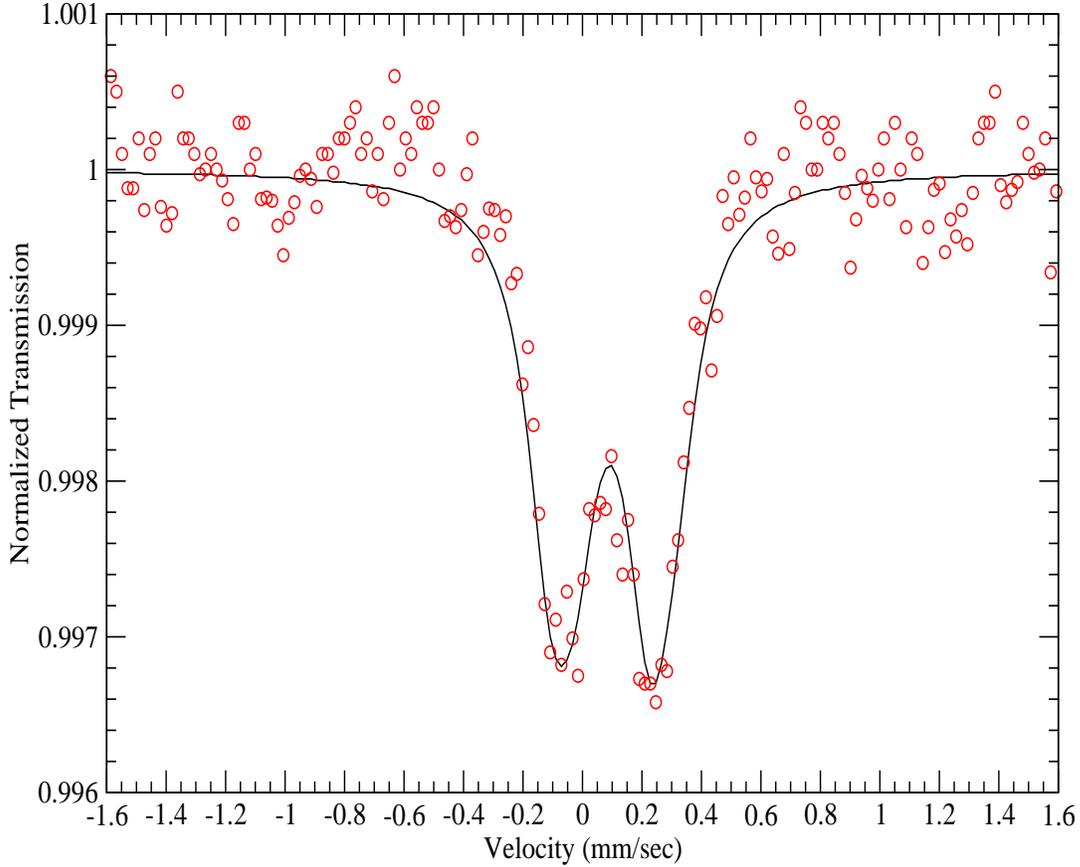}
\caption{(Color Online) $^{99}$Ru M\"{o}ssbauer spectra of CeRu$_2$ at 4.2 K(o).
The line is the spectrum determined by the calculated EFG's, Table 3,
using 0.09 mm sec$^{-1}$ as the  full-width half-maximum  and isomer shift equal to 0.09 msec$^{-1}$.
}
\label{CeRu2-4.2K}
\end{center}
\end{figure}

In Figure \ref{CeRu2-4.2K} we show the M\"{o}ssbauer spectrum of CeRu$_2$ at the Ru site at 4.2 K.
The  full line is  the spectrum determined with
values of the electric field gradient tensor (EFG) at the Ru site
in Table 3.
Other parameters in these spectra are the isomer shift, IS=0.12 mm/sec, and the half-width of lines, $\Gamma$=0.11 mm/sec.

The calculations discussed here have demonstrated that 
it is possible for transferred hyperfine magnetic fields at the Ru sites
to be very small in these ferromagnetic materials and allow us to reconcile the
apparent contradiction between the results of the $^{99}$Ru ME measurements
and those of the transport, magnetic susceptibility, magnetization, 
and neutron diffraction experiments on GdRu$_2$ and HoRu$_2$.
\noindent

\section{ Discussion}

A basic assumption of using the $^{99}$Ru ME is that the spectrum reflects the electronic
environment in the material and that magnetic order is reflected in an induced hyperfine field.
However the transferred hyperfine magnetic fields in ferromagnetic 
GdRu$_2$ and HoRu$_2$  are so  small that one would conclude that these materials do not have magnetic order.
The calculated properties of these materials have shown how this apparent discrepancy between M\"{o}ssbauer, 
the collapse of the hyperfine magnetic field,
 and results of neutron diffraction, magnetization, transport and specific 
measurements on the same samples  arise.
Analyzing the calculated contributions to $B_{hyperfine}$,
the net electronic spin, S, on the Gd and Ho in Tables 1 and 2 is roughly proportional to the magnitude of the
largest contributions, $B_{core}$ and $B_{valence}$.
This suggests that $B_{hyperfine}$ is $\simeq 0$ on the Ru
sites because S is $\simeq 0$ on these sites.
This arises because 4$d$ Ru electrons form polarized conduction bands rather than localized moments.
Consequently the $^{99}$Ru ME is misleading regarding magnetic order in the RRu$_2$ intermetallics. 

In contrast, moments in the magnetically ordered ruthenates are shared between 
Ru and O sites\cite{Singh1996,Mazin1997}
and the temperature dependence of magnetic order is reflected in $B_{hyperfine}$
determined from the M\"{o}ssbauer spectra.
The 2$p$ orbitals of the six neighboring oxygen atoms
in the RuO octahedra are strongly hybridized with Ru 4$d$ electrons.
A possibile explanation for the different behavior of Ru in the ruthenates and intermetallics
lies in the difference in the  electronegativities between Ru and O compared to that between Ru and rare earth atoms.
The Pauling electronegativity of oxygen atoms is 3.44,
which
is larger than that
of Ru(2.2)\cite{Mullally}.
In GdRu$_2$ and HoRu$_2$,  the opposite is the case. For Gd and Ho, the
electronegativity values are 1.1 to 1.25.
As a result the rare-earth atoms lose their 5$d$ and two 6$s$ electrons which take on the character of Ru $d$ electrons.
Competition for electrons between Ru atoms 
leads to the wide $d$-bands seen in the bandstucture.
 The structure can be pictured as a lattice of positive Gd and Ru ions,
with  the 4$f$ moments tightly-bound on the rare-earth sites
 as in the rare-earth metals\cite{Coqblin1977}.
The 4$d$ derived conduction bands 
mediate the coupling between the rare earth moments rather than forming
localized moments on the Ru sites.

There seem to be few other examples where hyperfine magnetic fields have not been induced by magnetic order.
There is evidence for a collapse of $B_{hyperfine}$ in 
hexagonal close-packed (hcp) Fe, which is the stable phase at high pressures.
Whereas there is evidence that this phase is antiferromagnetically ordered
from Raman scattering,  the six-line $^{57}$Fe M\"{o}ssbauer spectrum disappears at
the transition from bcc Fe to hcp Fe with increasing pressure\cite{Cort1982}.
$B_{hyperfine}$ was calculated by Seinle-Neumann et al.\cite{Neumann2004}
also using the wien2k software.  They found that there was almost complete
cancellation of the large core and 
valence contributions to B$_{con}$ as the atomic volume was reduced, 
simulating increasing pressure.
In determining the lattice constants for GdRu$_2$ at low temperature,
we also found that the sign of $B_{hyperfine}$ changed as the lattice constants were varied,
although not at the values of interest.
However, the contributions to B$_{hyperfine}$ in GdRu$_2$ and HoRu$_2$
are less than 6 T due to the itinerant nature of the 4$d$ electrons.

The results of this investigation point to the unsuspected sensitivity of a 
nuclear probe, such as the M\"{o}ssbauer Effect, to the details of the 
electronic structure in magnetically ordered materials.
The {\it ab initio} calculations provide a quantitative description of 
how the collapse of the hyperfine magnetic field at the Ru site arises.
In doing so, it was shown that the M\"{o}ssbauer spectrum can be a probe 
of bandstructure, as well as the local electronic environment.

\section{Acknowledgments}
Work was supported by the USDOE(DE-FG02-03ER46064), by CCSA$\#$7669 at 
CSU-Fresno, 
and by USDOE(DE-GF02-04ER46105), and NSF(DMR0802478), at UCSD.
D. C. wishes to thank Dr. M.  D. Jones of the Center for Computational Research 
at the University of Buffalo for his help in using the Wien2k package.

\end{document}